\documentclass[10pt,preprintnumbers,floatfix,letterpaper,onecolumn,aps,prd,epsfig,nofootinbib,longbibliography]{revtex4-1}

\usepackage{natbib}
\usepackage{appendix}
\usepackage{graphicx}
\usepackage{subcaption}
\usepackage[utf8]{inputenc}
\usepackage[T1]{fontenc}
\usepackage{physics}
\usepackage{latexsym}
\usepackage{epstopdf}
\usepackage{latexsym}
\usepackage{amssymb}
\usepackage{amsmath}
\usepackage{mathtools}
\usepackage{color}
\usepackage{mathrsfs}
\usepackage{xparse}
\usepackage{enumerate}
\usepackage{multirow,tabularx,boldline,array}
\usepackage{bbding}
\usepackage{diagbox}
\usepackage{siunitx}
\usepackage{enumitem}
\usepackage{pifont}
\usepackage{siunitx}
\usepackage[
pdfstartview=FitH,
bookmarksnumbered=true,
bookmarksopen=true,
colorlinks,
linkcolor=blue,
anchorcolor=green,
citecolor=blue
]{hyperref}
\usepackage{makecell}

\begin{document}
	
	\renewcommand\arraystretch{2}
	\newcommand{\bq}{\begin{equation}}
		\newcommand{\eq}{\end{equation}}
	\newcommand{\bqn}{\begin{eqnarray}}
		\newcommand{\eqn}{\end{eqnarray}}
	\newcommand{\nb}{\nonumber}
	\newcommand{\lb}{\label}
	\newcommand{\cb}{\color{blue}}
	\newcommand{\cc}{\color{cyan}}
	\NewDocumentCommand{\evalat}{sO{\big}mm}{%
		\IfBooleanTF{#1}
		{\mleft. #3 \mright|_{#4}}
		{#3#2|_{#4}}%
	}
	\newcommand{\PRL}{Phys. Rev. Lett.}
	\newcommand{\PL}{Phys. Lett.}
	\newcommand{\PR}{Phys. Rev.}
	\newcommand{\CQG}{Class. Quantum Grav.}
	

    \title{Gravitational lensing by charged black hole with global monopole in the strong field limit}

	\author{Yi-Ling Lan$^{1}$}
	\affiliation{	$^1$ Department of Physics and Synergetic Innovation Center for Quantum Effects and Applications, Hunan Normal University, 36 Lushan Road, Changsha, Hunan 410081, China\\
		$^2$ Institute of Interdisciplinary Studies, Hunan Normal University, 36 Lushan Road, Changsha, Hunan 410081, China}
	
	\author{Yun-Feng Qu$^{1}$}
	\affiliation{	$^1$ Department of Physics and Synergetic Innovation Center for Quantum Effects and Applications, Hunan Normal University, 36 Lushan Road, Changsha, Hunan 410081, China\\
		$^2$ Institute of Interdisciplinary Studies, Hunan Normal University, 36 Lushan Road, Changsha, Hunan 410081, China}
	
	\author{Jiawei Hu$^{1,2}$}
	\affiliation{	$^1$ Department of Physics and Synergetic Innovation Center for Quantum Effects and Applications, Hunan Normal University, 36 Lushan Road, Changsha, Hunan 410081, China\\
		$^2$ Institute of Interdisciplinary Studies, Hunan Normal University, 36 Lushan Road, Changsha, Hunan 410081, China}
	
	\author{Hongwei Yu$^{1,2}$}
	\email[Contact author: ]{hwyu@hunnu.edu.cn}
	\affiliation{	$^1$ Department of Physics and Synergetic Innovation Center for Quantum Effects and Applications, Hunan Normal University, 36 Lushan Road, Changsha, Hunan 410081, China\\
		$^2$ Institute of Interdisciplinary Studies, Hunan Normal University, 36 Lushan Road, Changsha, Hunan 410081, China}

	
	\begin{abstract}
		
	We investigate gravitational lensing near a charged black hole with a global monopole in the strong field regime, focusing on  the combined effects of the global monopole and black hole charge on key observables in gravitational lensing   both analytically and numerically.   Our results reveal that the dependence of the angular separation on charge is intricately tied to the deficit angle caused by the global monopole. In particular, we identify three critical values of the global monopole parameter that determine whether the angular separation increases monotonically, decreases monotonically, or exhibits extrema as the charge varies. A similar complex dependence is found for the flux ratio as a function of the deficit angle, and for the magnification of the first relativistic image as a function of charge.
    These behaviors contrast sharply with the monotonic changes observed in the absence of either a global monopole or charge.    Our findings highlight that the effects of the charge and global monopole on gravitational lensing cannot be described as simple additive contributions. Instead, their combined effects lead to a rich and interdependent behavior that enhances our understanding of strong-field gravitational lensing.     While the charge and global monopole are expected to be small in typical astrophysical contexts,     the results presented here could be experimentally explored in analog gravity systems, where these parameters are not constrained. This opens the door to potential experimental verification of the phenomena predicted in this study.

    \end{abstract}

	\maketitle

	\section{Introduction}
	\renewcommand{\theequation}{1.\arabic{equation}}\setcounter{equation}{0}

	Gravitational lensing is a phenomenon in which the path of light from a distant object is bent by the gravitational field of a massive intervening object (referred to as the lens).  This lens could be a black hole, a galaxy, or a galaxy cluster positioned between the source and the observer.  The bending of light due to gravity can produce various observable effects, including multiple images of the same object, magnification of the source's brightness, and the formation of distinctive ringlike structures around the lens, known as Einstein rings. Originally predicted by Albert Einstein within the framework of general relativity \cite{Einstein:1936llh},  gravitational lensing has since been extensively observed and confirmed through astronomical observations \cite{Walsh:1979nx}.

    Gravitational lensing is fundamentally characterized by the deflection angle, which quantifies the bending of light due to gravity. Although the deflection angle itself is not directly observable, it  serves as the basis for calculating key observables, such as the angular separation between the first and subsequent images, the ratio of the flux intensity of the first image to the cumulative flux intensity of all other images, the magnification of the images, and the size of the Einstein rings.
    Accurately determining the deflection angle is therefore essential for a comprehensive understanding of gravitational lensing.  However, in general situations, directly calculating the deflection angle is challenging.    
    Fortunately, in regimes where the gravitational field is either very weak or extremely strong, approximate methods can be employed. 
    In the weak field limit, where the deflection is small, linearized general relativity provides reliable estimates.  Conversely, in strong-field environments, such as near black holes or neutron stars, the gravitational field is strong enough to cause substantial bending of light, sometimes even allowing light to orbit the lens multiple times. In these cases, numerical methods or specialized analytical approximations tailored to highly curved spacetimes are necessary.
     Here let us note that, Virbhadra and Ellis have developed a gravitational lens equation that allows for arbitrarily large deflection angles, which has been applied to analyze gravitational lensing by a Schwarzschild black hole using numerical methods \cite{Virbhadra:1999nm}. 
    Based on this equation, the position and magnification of relativistic images for a Schwarzschild black hole were derived analytically in Ref. \cite{Bozza:2001xd}. 
    Gravitational lensing in the strong field limit for a general spherically symmetric and static spacetime has been investigated by Bozza \cite{Bozza:2002zj} and later by Tsukamoto \cite{Tsukamoto:2016jzh}. In particular, the  Reissner-Nordstr\"{o}m (RN) black holes have been studied as one of the specific examples in Refs. \cite{Bozza:2002zj,Tsukamoto:2016jzh}. Subsequent studies have explored gravitational lensing in Kerr black holes \cite{Bozza:2002af,Bozza:2006nm,Bozza:2005tg}, Kerr-Newman black holes \cite{Hsieh:2021scb,Chen:2024oyv}, 
    as well as various other curved spacetimes \cite{Eiroa:2004gh,Whisker:2004gq,Tsukamoto:2012xs,Tsukamoto:2016qro,Gyulchev:2006zg,Tsukamoto:2020bjm,Tsukamoto:2021caq,Chagoya:2020bqz,Younas:2015sva,Vachher:2024ldc,Virbhadra:2002ju,Soares:2023uup}.  
    These studies highlight the role of spacetime curvature in determining lensing phenomena.

    Apart from spacetime curvature, the topology of spacetime can also influence gravitational lensing \cite{Cheng:2010nd,Man:2012ivp,Sharif:2015kna,Soares:2024rhp}. A notable example is a spacetime with a global monopole, a topological defect believed to be formed during phase transitions in the early universe \cite{Barriola:1989hx}. Global monopoles, which arise from phase transitions in systems with self-coupled triplet scalar fields, have garnered significant interest in theoretical physics. Their formation occurs when the original global $O(3)$ symmetry is spontaneously broken to $U(1)$, resulting in a topological defect. These monopoles introduce a deficit solid angle in the surrounding spacetime, leading to distinctive  effects,  including   the bending of light \cite{Barriola:1989hx} and black hole 
    thermodynamics \cite{Yu:1994}.     For instance, a Schwarzschild black hole that absorbs a global monopole exhibits,  in the strong field limit, a reduced relative light flux in its primary image while simultaneously increasing the angular separation between the primary image and the subsequent images \cite{Cheng:2010nd}.
    While previous studies have focused on gravitational lensing effects for black holes with either a global monopole \cite{Cheng:2010nd} or with charge \cite{Bozza:2002zj,Tsukamoto:2016jzh} individually, the combined influence of both charge and a global monopole on gravitational lensing remains unexplored. This raises an intriguing question: Are the effects of a global monopole and charge on black hole gravitational lensing independent, or do they interact in nontrivial ways?
	
	In this paper, we investigate gravitational lensing by a charged black hole with a global monopole in the strong field limit. In this regime, photons can orbit the lens multiple times, producing a series of relativistic images. We analyze key observables associated with these images, including the angular separation and flux ratio between the first and subsequent images, the angular size of the first Einstein ring, the magnification of the first image, and the time delay between the first and second images.
   Our analysis reveals intricate and nontrivial dependencies of these observables on the black hole charge and the global monopole parameter. In particular,  the angular separation exhibits diverse behaviors as the charge increases, depending on the relationship between the global monopole parameter and three critical values. Specifically, the angular separation may decrease monotonically, increase monotonically, initially decrease and then increase, or even undergo a more complex pattern of initial decrease, subsequent increase, and another decrease at higher charge values. 
    Similarly, a critical charge exists that determines the behavior of the flux ratio with respect to the global monopole parameter. When the charge is below this critical value, the flux ratio increases monotonically with the global monopole parameter, whereas  for charge values  exceeding  the critical value, the flux ratio initially decreases and then increases.  Additionally, we identify two critical values of the global monopole parameter that determine whether the magnification of the first relativistic image monotonically decreases, increases, or exhibits nonmonotonic dependence on the charge.
    These rich behaviors stand in stark contrast to the simpler, monotonic trends observed in the absence of either charge or a global monopole.  Our results highlight that the effects of charge and the deficit solid angle due to the presence of the global monopole are not independent but rather interact in a nontrivial manner, significantly altering the gravitational lensing characteristics. 
    Furthermore, we numerically calculate the observables for scenarios in which the central black holes in the Milky Way, M87, and M104 galaxies act as gravitational lenses,  comparing configurations with and without charge and a global monopole.   While the resulting differences are small and likely challenging to detect in astrophysical contexts,   the growing interest in gravitational analogs \cite{Barcelo:2005fc} offers alternative experimental settings. In such analog systems, effective charge and monopolelike parameters need not be small, making our predicted effects potentially observable in laboratory-based analog gravity experiments.

    The paper is organized as follows. In Sec. \ref{apply}, we derive the deflection angle for an RN black hole with a global monopole in the strong field limit. 
    In Sec. \ref{sec3}, we analyze the effects of charge and global monopole on typical  gravitational lensing observables both analytically and numerically. 
    In Sec. \ref{Discussion}, we discuss the potential detectability of these effects in astrophysical settings and their implications in analog gravity systems. 
    Finally, we summarize in Sec. \ref{conclusion}. Throughout this paper, we adopt natural units $\hbar=c=G_N=1$, where $\hbar$ is the reduced Planck constant, $c$ is the speed of light, and $G_N$ is the Newtonian constant of gravitation.

	 \section{Deflection angle in strong field limit} \lb{apply}
	\renewcommand{\theequation}{2.\arabic{equation}}\setcounter{equation}{0}

    We now investigate the gravitational lensing of an RN black hole with a global monopole. Its line element is given by \cite{Barriola:1989hx,Guendelman:1991qb}
	\bq \lb{eq:line}
	ds^2=-A(r)dt^2+B(r)dr^2+C(r)\left(d\theta ^2+\sin^2\theta d\varphi^2\right),
	\eq
	where 
	\bq
	A(r)=B^{-1}(r)=1-8\pi\gamma^2-\frac{2M}{r}+\frac{Q^2}{r^2},\quad C(r)=r^2.
	\eq
    Here, $M$ and $Q$ represent the mass and charge of the black hole, respectively, and $\gamma$ is a parameter associated with the scale of gauge-symmetry breaking.  
The presence of a global monopole induces a deficit solid angle in spacetime, quantified by $\Delta=32\pi^2\gamma^2$ \cite{Barriola:1989hx}. This deficit angle modifies the geometry of spacetime, impacting the gravitational lensing properties.

For convenience, we introduce two dimensionless parameters to characterize the effects of the global monopole and charge:
We define the global monopole parameter $\epsilon$ as   
\bq
\epsilon=1-8\pi\gamma^2=1-\frac{\Delta}{4\pi},
\eq
which ranges from $\epsilon = 0$ (corresponding to a maximum deficit solid angle $\Delta=4\pi$)  to $\epsilon = 1$ (when no monopole is present, i.e., $\Delta=0$), 
and the charge  parameter $\eta$ as 
\begin{equation}
    \eta=Q^2/M^2.
\end{equation}
This parameter quantifies the effect of charge, with values ranging from $\eta=0$ (uncharged black hole) to $\eta=1$ (extremal black hole in standard RN spacetime).
To ensure that the spacetime does not develop a naked singularity, we impose the constraint $|Q| \leq M/\sqrt{\epsilon}$. For generality, and to avoid naked singularities for any $\epsilon$, we restrict the charge to $|Q|\leq M$, leading to the constraint $\eta \in [0,1]$. With these dimensionless parameters, the functions $A(r)$ and $B(r)$ in the metric can be rewritten as
\begin{equation}\label{ABC-dless}
    A(r)=B^{-1}(r)=\epsilon-\frac{2M}{r}+\frac{\eta M^2}{r^2}.
\end{equation}
In the following, we calculate the deflection angle in the strong field limit. We start by introducing the relevant physical quantities associated with gravitational lensing and provide a precise definition of the strong field regime.

When a photon is emitted from a distant light source, it undergoes gravitational deflection as it passes near a lensing celestial body. The {\it impact parameter} $b$ is defined as the shortest distance between the straight-line path of a photon if it were unaffected by gravity and the center of the lensing object. In addition, there exists a closest distance $r_0$ between the photon and the center of the lensing celestial body, and the point $r=r_0$ is commonly referred to as the {\it turning point}. Based on the conservation of energy and angular momentum, the impact parameter $b$ and the turning point $r_0$ are related by the equation
	\bq
	b=\sqrt{\frac{C\left( r_0 \right)}{A\left( r_0 \right)}}.
	\eq
 In the vicinity of the event horizon of a black hole, there exists a particularly interesting region known as the {\it photon sphere}. On the photon sphere, photons can orbit the black hole along circular trajectories. The radius of the photon sphere, usually denoted as $r_m$, can be determined by solving the following equation \cite{Claudel:2000yi}:
	\bq\lb{eq:dV_eff}
	\frac{C'(r_m)}{C(r_m)}-\frac{A'(r_m)}{A(r_m)}=0,
	\eq
	where $'$ denotes differentiation with respect to the radial coordinate $r$. This circular orbit is unstable, which means that any small perturbation can cause photons to either fall into the black hole or escape to infinity. Consequently, $r_m$ is the closest distance between observable photons and black hole.
	The photon sphere also defines the critical impact parameter as \cite{Bozza:2002zj,Tsukamoto:2016jzh}
	\bq\lb{eq:b_m}
	b_m=\lim_{r_0\rightarrow r_m}\sqrt{\frac{C(r_0)}{A(r_0)}}.
	\eq
    For an RN black hole with a global monopole, the radius of the photon sphere and the critical impact parameter are given by
    \bq
    r_m\left(\epsilon ,\eta \right)
    =\frac{3+\sqrt{9-8\epsilon \eta}}{2\epsilon}M,
    \eq
    and
    \bq\lb{eq:b_m}
	b_m\left(\epsilon ,\eta \right)=\frac{\bar{r}_m^{2}}{\sqrt{\epsilon \bar{r}_m^{2} -2\bar{r}_m +\eta}}M,
	\eq
    respectively, where $\bar{r}_m\left(\epsilon ,\eta \right)=r_m\left(\epsilon ,\eta \right)/M=(3+\sqrt{9-8\epsilon \eta})/2\epsilon$ is the dimensionless radius of the photon sphere. 
       The limit $r_0\rightarrow r_m$ or $b\rightarrow b_m$ is referred to as the {\it strong deflection limit}, also known as the {\it strong field limit}. Meanwhile, the parameter $b_m$ is the smallest impact parameter for observable photons. If the impact parameter of a photon is less than $b_m$, it will inevitably fall into the black hole and cannot be observed.
    
    Generally, for a photon originating from a source at infinity, undergoing deflection at the turning point $r=r_0$, and then escaping toward a distant observer, the deflection angle $\alpha(r_0)$ can be derived as
    \bq\lb{eq:alpha}
	 \alpha \left( r_0 \right) =I\left( r_0 \right) -\pi,
	\eq
	where
	\bq
	 I\left( r_0 \right) =2\int_{r_0}^{\infty}{\frac{dr}{\sqrt{\frac{C(r)R(r)}{B(r)}}}},
	\eq
	with
	\bq
	 R(r)= \frac{A(r_0)C(r)}{A(r)C(r_0)}-1.
	\eq
    
    In the strong field limit, photons undergo multiple orbits near the photon sphere, causing the deflection angle $\alpha$ to diverge. Specifically, $I(r_0)$ diverges as $r_0\rightarrow r_m$. To address this issue, a new variable $z$ is introduced, which is defined as \cite{Tsukamoto:2016jzh}
    \bq\lb{z}
	z=1-\frac{r_0}{r}.
    \eq
    Then, $I(r_0)$ can be expressed as
	\bq\lb{eq:I_r0}
	I\left( r_0 \right) =\int_0^1{f\left( z,r_0 \right) dz},
	\eq
	where
    \bq
    f(z,r_0)=\frac{2r_0\sqrt{B(z,r_0)}}{\sqrt{R(z,r_0)C(z,r_0)(1-z)^4}}.
    \eq
   Here, $B(z,r_0)$, $C(z,r_0)$, and $R(z,r_0)$ are obtained by applying the variable substitution in Eq. \eqref{z} to the original expressions for $B(r)$, $C(r)$, and $R(r)$, respectively. 

   To isolate the divergent part of the integral \eqref{eq:I_r0}, we partition the integral $I(r_0)$ into a divergent part $I_D(r_0)$ and a regular part $I_R(r_0)$ as
\bq\lb{eq:I}
  I(r_0)=I_D(r_0)+I_R(r_0).
   \eq
   Specifically, the divergent part $I_D(r_0)$ is defined as
   \bq\label{eq:ID1}
  I_{D}(r_{0})= \int^{1}_{0}f_{D}(z,r_{0})dz,
  \eq
  where
  \bq\label{eq:fD1}
  f_{D}(z,r_{0})= \frac{2r_{0}}{\sqrt{c_{1}(r_{0})z+c_{2}(r_{0})z^{2}}},
  \eq
  and coefficients $c_{1}(r_{0})$ and $c_{2}(r_{0})$ are given by \cite{Tsukamoto:2016jzh}
  \bq\label{eq:c10}
  c_{1}(r_{0})=\frac{C(r_0)r_{0}}{B(r_0)}\left(\frac{C'(r_0)}{C(r_0)}-\frac{A'(r_0) }{A(r_0)}\right)
  \eq
and
  \bq\label{eq:c20}
  c_{2}(r_{0})=\frac{C(r_0)r_{0}}{B(r_0)} \left\{ \left(\frac{C'(r_0)}{C(r_0)}-\frac{A'(r_0) }{A(r_0)}\right) \left[  \left(\frac{C'(r_0)}{C(r_0)}-\frac{A'(r_0) }{A(r_0) }-\frac{B'(r_0)}{B(r_0)} \right)r_{0}-3 \right]  +\frac{r_{0}}{2} \left( \frac{C^{''}(r_0)}{C(r_0)}-\frac{A^{''}(r_0)}{A(r_0)} \right) \right\}.
  \eq
   Meanwhile, the regular part $I_{R}(r_{0})$ is given by
  \bq\label{eq:IR}
  I_{R}(r_{0})= \int^{1}_{0} \left[f(z,r_{0})-f_D(z,r_0)\right]dz.
  \eq
  After some calculations, the deflection angle $\alpha$ in the strong field limit is derived as \cite{Tsukamoto:2016jzh}
\bq\label{eq:angleb}
\alpha(b)=-\bar{a}\ln \left( \frac{b}{b_{m}}-1 \right) +\bar{b} +\mathcal{O}((b-b_{m})\ln(b-b_{m})),
\eq
where the coefficients $\bar{a}$ and $\bar{b}$ are given by
   \bq\label{eq:abar}
   \bar{a}=\sqrt{\frac{2B_mA_m}{C^{''}_{m}A_{m}-C_{m}A^{''}_{m}}},
   \eq
   and	
   \bq\label{eq:bbar}
   \bar{b}=\bar{a}\ln \left[r^{2}_{m}\left(\frac{C_{m}^{''}}{C_{m}}-\frac{A_{m}^{''}}{A_{m}}\right)\right] +I_{R}(r_{m})-\pi,
   \eq
   respectively, with the subscript $m$ denoting quantities evaluated at at $r=r_m$.  
    For a detailed derivation, see Ref. \cite{Tsukamoto:2016jzh}.

Substituting  the metric functions for an RN black hole with a global monopole Eq. \eqref{ABC-dless} into Eqs. \eqref{eq:abar} and \eqref{eq:bbar}, we obtain the coefficients $\bar{a}$ and $\bar{b}$ which determine the deflection angle in the strong field limit as
\bq\label{eq:abar_rad}
 \bar{a}(\epsilon,\eta)
 =\frac{\bar{r}_m\left( \epsilon ,\eta \right)}{\sqrt{\epsilon \bar{r}_m^{2}\left( \epsilon ,\eta \right) -2\eta}},
 \eq
and 
\bq\label{eq:bbar_rad}
\bar{b}(\epsilon,\eta)=\bar{a}(\epsilon,\eta)\ln \left[ \frac{8\left(3\bar{r}_m(\epsilon,\eta)-4\eta \right) ^3}{\bar{r}_m^{2}(\epsilon,\eta)\left(\bar{r}_m(\epsilon,\eta)-\eta \right) ^2}\left( 2\sqrt{\bar{r}_m(\epsilon,\eta)-\eta}-\sqrt{3\bar{r}_m(\epsilon,\eta)-4\eta} \right) ^2 \right] -\pi ,
\eq
respectively.
When the global monopole is absent  ($\epsilon=1$), the coefficients $\bar{a}$ and $\bar{b}$   recover the results for the RN black hole, as derived in Ref. \cite{Tsukamoto:2016jzh}. Similarly, in the absence of charge ($\eta=0$), the coefficients $\bar{a}$ and $\bar{b}$  reduce to the expressions obtained for a Schwarzschild black hole with a global monopole, as given in Ref. \cite{Cheng:2010nd}. Furthermore, in the absence of both charge and the global monopole ($\eta=0$, $\epsilon=1$), the coefficients $\bar{a}$ and $\bar{b}$ match the results from Refs. \cite{Bozza:2002zj,Tsukamoto:2016jzh} for a Schwarzschild black hole. These consistency checks further validate our approach, ensuring that the inclusion of both charge and a global monopole provides a unified framework that naturally extends existing results.

    \section{Interdependence of gravitational lensing observables on charge and global monopole}\label{sec3}
    \renewcommand{\theequation}{3.\arabic{equation}}\setcounter{equation}{0}

    While the deflection angle of light $\alpha$ and its associated coefficients $\bar{a}$ and $\bar{b}$ are crucial for understanding gravitational lensing, they are not directly measurable. To relate these theoretical parameters to observable quantities, the lens equation serves as a critical tool, converting these quantities into experimentally accessible ones.

\begin{figure}[h!]
		\includegraphics[height=6.5cm]{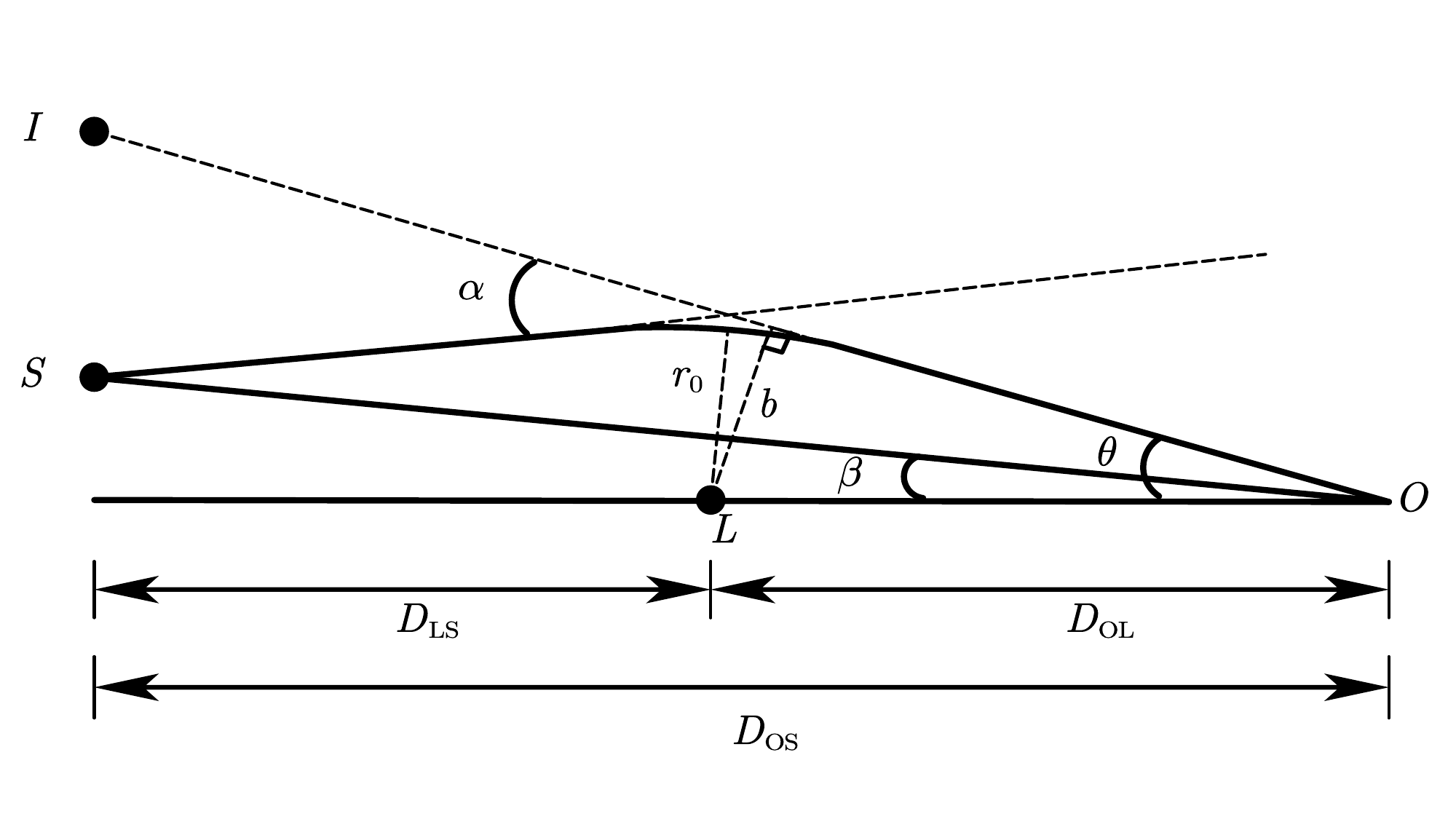}
		\caption{
			Fundamental geometric configuration underlying the phenomenon of gravitational lensing. $b$ is the impact parameter, $r_0$ represents the minimum distance between the light ray and the lens, and $I$ and $L$ denote the positions of the image and the lens. $S$ and $O$ represent the source and observer.
		}\label{fig:lensing}
	\end{figure}

	Consider the following scenario: A photon emanating from a celestial source ($S$) passes through the gravitational field of a foreground object, termed the lens ($L$). The light ray undergoes deflection, altering its original trajectory and reaching the observer ($O$)  at an angle $\theta$, which deviates from the angle $\beta$ that would have been observed in the absence of the lens.  The total deflection angle of the light is denoted by $\alpha$.  
	The lens equation can be formulated as follows \cite{Virbhadra:1999nm}:
	\bq \lb{eq:lens equation}
	\tan \beta =\tan \theta -\frac{D_{\text{LS}}}{D_{\text{OS}}}\left[ \tan \theta +\tan \left( \alpha -\theta \right) \right], 
	\eq
	where $D_\text{LS}$ represents the distance between the lens and the projection of the source onto the optical axis (OL). Meanwhile, $D_\text{OS}=D_\text{OL}+D_\text{LS}$, where $D_\text{OL}$ refers to the distance between the lens and the observer. Specifically, the optical path diagram of the gravitational lens is presented in Fig. \ref{fig:lensing}.     

The image formed by photons that encircle a lens $n$ times is referred to as the $n$th relativistic image. 
Using the lens equation \eqref{eq:lens equation} and the analytical expression for the deflection angle \eqref{eq:angleb}, the angular position of the $n$th relativistic image in the strong field regime is given by \cite{Bozza:2002zj}
\bq\lb{eq:theta_n}
	\theta_n=\theta^0_n+\frac{b_m e_n (\beta-\theta^0_n) D_{\text{OS}}}{\bar{a} D_{\text{LS}}D_{\text{OL}}},
   	\eq
where
\bq\label{theta-0-n}
	\theta^0_n=\frac{b_m}{D_\text{OL}}(1+e_n),
    \eq
    and 
    \begin{equation}\lb{eq:en}
        e_n=e^{\frac{\bar{b}-2n\pi}{\bar{a}}}.
    \end{equation}
    The magnification of the $n$th relativistic image is then given by \cite{Bozza:2002zj}
    \begin{equation}\lb{eq:mu}
        \mu_n = \left( \frac{\beta}{\theta} \frac{d\beta}{d\theta} \right)^{-1} \bigg|_{\theta_n^0} 
        =e_n\frac{b_{m}^{2}\left( 1+e_n \right) D_{\text{OS}}}{\bar{a}\beta D_{\text{OL}}^{2}D_{\text{LS}}},
    \end{equation}   
    When the source, lens, and observer are nearly perfectly aligned, i.e., $\beta=0$,  the angular radius of the $n$th Einstein ring is given by \cite{Bozza:2002zj}
    \bq\lb{eq:ring}
    \theta _{n}^{E}=\left( 1-\frac{D_{\text{OS}}}{D_{\text{LS}}D_{\text{OL}}}\frac{b_me_n}{\bar{a}} \right) \theta^0_n.
    \eq   
Since $b_m\ll D_\text{OL}$, the size of the Einstein ring can be well approximated by $\theta^0_n$ as given in Eq. \eqref{theta-0-n}.

    In addition to the observables discussed above, there are three other essential observables in gravitational lensing.
    
  First, there is the angular separation.     	
	It is hypothesized that the outermost (first) image in the series of images stands apart from the others, being uniquely discernible, while the remaining images nearly overlap and are difficult to distinguish. Therefore, we introduce an observable quantity: the angular separation, denoted as $s$, which quantifies the difference between the angular position of the first image and that of the subsequent ones. This quantity is defined as \cite{Bozza:2002zj}
	\bq\lb{eq:s}
	s=\theta_1-\theta_\infty\approx
        \frac{b_m}{D_\text{OL}}e_1.
	\eq

    Second, there is the flux ratio. The flux ratio, also known as the relative magnification, quantifies the brightness contrast between the primary image and the cumulative flux of all subsequent relativistic images.
    It is given by  \cite{Bozza:2002zj}
	\bq
	\mathcal{R}=\frac{\mu_1}{ {\textstyle \sum_{n=2} ^{\infty}} \mu_n},
	\eq
	where
	\bq
	\sum_{n=2}^{\infty} \mu_{n}=\frac{b_{m}^{2}D_{OS}}{\bar{a}\beta D_{OL}^{2}D_{LS}}\frac{e^{\frac{2\bar{b}-4\pi}{\bar{a}}}+e^{\frac{\bar{b}-2\pi}{\bar{a}}}+e^{\frac{\bar{b}}{\bar{a}}}}{e^{\frac{4\pi}{\bar{a}}}-1}.
	\eq
	After simplification, the flux ratio becomes
	\bq\lb{eq:flux}
	\mathcal{R}=\frac{(e^{\bar{b}/\bar{a}}+e^{2\pi / \bar{a}})(e^{4\pi /\bar{a}}-1)}{e^{\bar{b}/\bar{a}}+e^{2\pi/\bar{a}}+e^{4\pi/\bar{a}}}.
	\eq
	    
     Third, there is the time delay. In the strong field regime, photons can orbit a black hole multiple times, leading to the formation of multiple relativistic images. Due to the differences in the path lengths taken by these photons, the corresponding light travel times are also different. The time delay between the $n$th and $k$th relativistic images, which are formed by photons traveling along different paths on the same side of the lensing object, is given by \cite{Bozza:2003cp}
    \bq\lb{eq:timedelay}
    \Delta T_{n,k}^{s}=b_m\left[ 2\pi \left( n-k \right) +2\sqrt{2}\left( \sqrt{e_k}-\sqrt{e_n} \right) \pm \frac{\sqrt{2}D_{\text{OS}}\beta}{\bar{a}D_{\text{LS}}}\left( \sqrt{e_k}-\sqrt{e_n} \right) \right],
    \eq
    where the positive (negative) sign applies when both images are located on the same (opposite) side of the source.
    For images on opposite sides of the lens, the time delay is given by \cite{Bozza:2003cp}
    \bq
    \Delta T_{n,k}^{o}=b_m\left[ 2\pi \left( n-k \right) +2\sqrt{2}\left( \sqrt{e_k}-\sqrt{e_n} \right) +\frac{\sqrt{2}D_{\text{OS}}\beta}{\bar{a}D_{\text{LS}}}\left( \sqrt{e_k}-\sqrt{e_n} \right) -\frac{2D_{\text{OS}}\beta}{D_{\text{LS}}} \right].
    \eq

By leveraging the coefficients $\bar{a}$ and $\bar{b}$ of  the deflection angle,  we now derive explicit analytical expressions for the key observables in gravitational lensing, including the angular position and the magnification of the $n$th relativistic image, the angular radius of the Einstein ring, the angular separation, the flux ratio, and the time delay. 
Previous studies have primarily relied on numerical approaches to analyze these observables in specific scenarios. For instance: Reference \cite{Bozza:2002zj} conducted numerical computations for the angular separation $s$  and the flux ratio $\mathcal{R}$ of an RN black hole.  
Reference \cite{Cheng:2010nd} focused on a Schwarzschild black hole with a global monopole, again without explicit analytical expressions for the observables. In this work, we go beyond previous studies by deriving explicit analytical expressions for both $s$  and  $\mathcal{R}$ for an RN black hole with a global monopole, facilitating a direct evaluation of how these observables respond to variations in charge ($\eta$) and the global monopole parameter ($\epsilon$).
To achieve a comprehensive understanding, we conduct both analytical and numerical analyses, examining both the individual effects of the global monopole and charge on the observables, as well as  the combined impact of charge and the global monopole, revealing their interplay in modifying strong field lensing behavior.

\subsection{Angular separation}

The angular separation between the first image and the subsequent images can be obtained using Eq. \eqref{eq:s} as
\begin{align}\lb{eq:s_rad}
s(\epsilon,\eta)=&\frac{M}{D_{\text{OL}}\sqrt{\epsilon \bar{r}_m^{2}-2\bar{r}_m+\eta}} \left[\frac{8\left( 3\bar{r}_m-4\eta \right) ^3}{\left( \bar{r}_m-\eta \right) ^2}\right]\left( 2\sqrt{\bar{r}_m-\eta}-\sqrt{3\bar{r}_m-4\eta} \right) ^2  \exp \left( \frac{-3\pi \sqrt{\epsilon \bar{r}_m^{2} -2\eta}}{\bar{r}_m} \right) ,
\end{align}
where $\bar{r}_m$ represents $\bar{r}_m(\epsilon,\eta)$ for brevity.

Firstly, we examine the dependence of the angular separation on the global monopole parameter for a fixed charge.  When the charge vanishes ($\eta=0$), the angular separation is given by
\bq\lb{eq:s_0}
s(\epsilon,0)=\frac{648\sqrt{3}\left( 2-\sqrt{3} \right) ^2M}{D_{\text{OL}} \epsilon^{3/2}}e^{ -3\pi \sqrt{\epsilon}}.
\eq
It is evident that the angular separation $s$ increases as the global monopole parameter $\epsilon$ decreases (i.e., as the deficit solid angle $\Delta$ caused by the global monopole  increases). This means that the presence of a global monopole results in a larger angular separation between the first image and the subsequent images, which is consistent with the numerical results in Ref. \cite{Cheng:2010nd}. 
When the charge is nonzero, the analytical expression becomes more complex. However, it is direct to check numerically that the partial derivative of the angular separation $s$ with respect to the global monopole parameter $\epsilon$ in the region $\epsilon\in [0,1]$ and $\eta\in [0,1]$ is always negative, meaning that the angular separation always increases as the deficit solid angle $\Delta$ caused by the global monopole increases, regardless of whether the black hole is charged or not.

Secondly, we investigate the dependence of the angular separation on the charge for a fixed global monopole parameter. First, when there is charge but no global monopole  (i.e., $\epsilon=1$), the angular separation is given by the expression
\bq\lb{eq:s_0_v_rad}
s(1,\eta)=\frac{M}{D_{\text{OL}}\sqrt{ w^{2}-2w+\eta}} \left[\frac{8\left( 3w-4\eta \right) ^3}{\left( w-\eta \right) ^2}\right]\left( 2\sqrt{w-\eta}-\sqrt{3w-4\eta} \right) ^2  \exp \left( \frac{-3\pi \sqrt{ w^{2} -2\eta}}{w} \right),
\eq
where $w$  represents the dimensionless radius of the photon sphere, $\bar{r}_m(1,\eta)$.  Furthermore, as shown in Appendix~\ref{ds}, the derivative of the angular separation with respect to $\eta$ is always positive. This indicates that the angular separation monotonically increases with increasing charge. 
Now, we analyze the behavior of the angular separation $s$ in the presence of both charge and a global monopole. Directly analyzing Eq. \eqref{eq:s_rad} is challenging due to its complexity. To understand the effects of both charge and the global monopole, we expand the angular separation expression for small and large charge values, respectively. 
For small charge ($\eta\ll1$), Eq. \eqref{eq:s_rad} can be expanded as
\bq \lb{eq:expand1}
s=\frac{648\left( 7\sqrt{3}-12 \right) e^{-3\pi \sqrt{\epsilon}}M}{\epsilon ^{3/2}D_{OL}}+\frac{108\left[ 2\pi \left( 7\sqrt{3}-12 \right) \sqrt{\epsilon}-51\sqrt{3}+88 \right]e^{-3\pi \sqrt{\epsilon}}M}{\sqrt{\epsilon}D_{OL}}\eta +\mathcal{O}\left( \eta^2 \right). 
\eq
 
The dependence of the angular separation on charge is determined by the coefficient of the $\eta$-dependent term. There exists a critical value of the global monopole parameter, given by $\epsilon _0=\left[ \frac{51\sqrt{3}-88}{2\pi \left( 7\sqrt{3}-12 \right)} \right] ^2\approx 0.1833742$, which separates two behaviors: 
For $\epsilon > \epsilon _0$, coefficient of the $\eta$-dependent term is positive, and the angular separation increases with charge. Conversely, for $\epsilon < \epsilon _0$, the coefficient is negative, and the angular separation decreases with charge. 
For large charge ($\eta\to 1$), the angular separation  (Eq. \eqref{eq:s_rad})  behaves as
\bq \lb{eq:se2}
s=H\left( \epsilon \right) +F\left( \epsilon \right) \left( 1-\eta \right) +\mathcal{O}\left( (1-\eta)^2 \right), 
\eq
where $H(\epsilon)$ and $F(\epsilon)$ are functions of the global monopole parameter $\epsilon$. Due to the complexity of these expressions, their explicit forms are provided in Appendix \ref{appendix}. We find that, similar to the small-charge case, there exists a critical value $\epsilon _1 \approx 0.1826688$, 
above which $F(\epsilon)$ is negative and the angular separation increases with charge, and below which $F(\epsilon)$ is positive and the angular separation decreases with charge. 
The analysis reveals that the global monopole parameter introduces a nonmonotonic behavior in the angular separation with respect to charge. There are critical values of 
$\epsilon$
 ($\epsilon_0$ and $\epsilon_1$) that determine whether the angular separation increases or decreases with charge. This behavior is distinctly different from the RN black hole case, where the angular separation always increases with charge.

\begin{figure}[htbp]
	\centering
    \begin{subfigure}[b]{0.3\textwidth}
	\includegraphics[height=5cm]{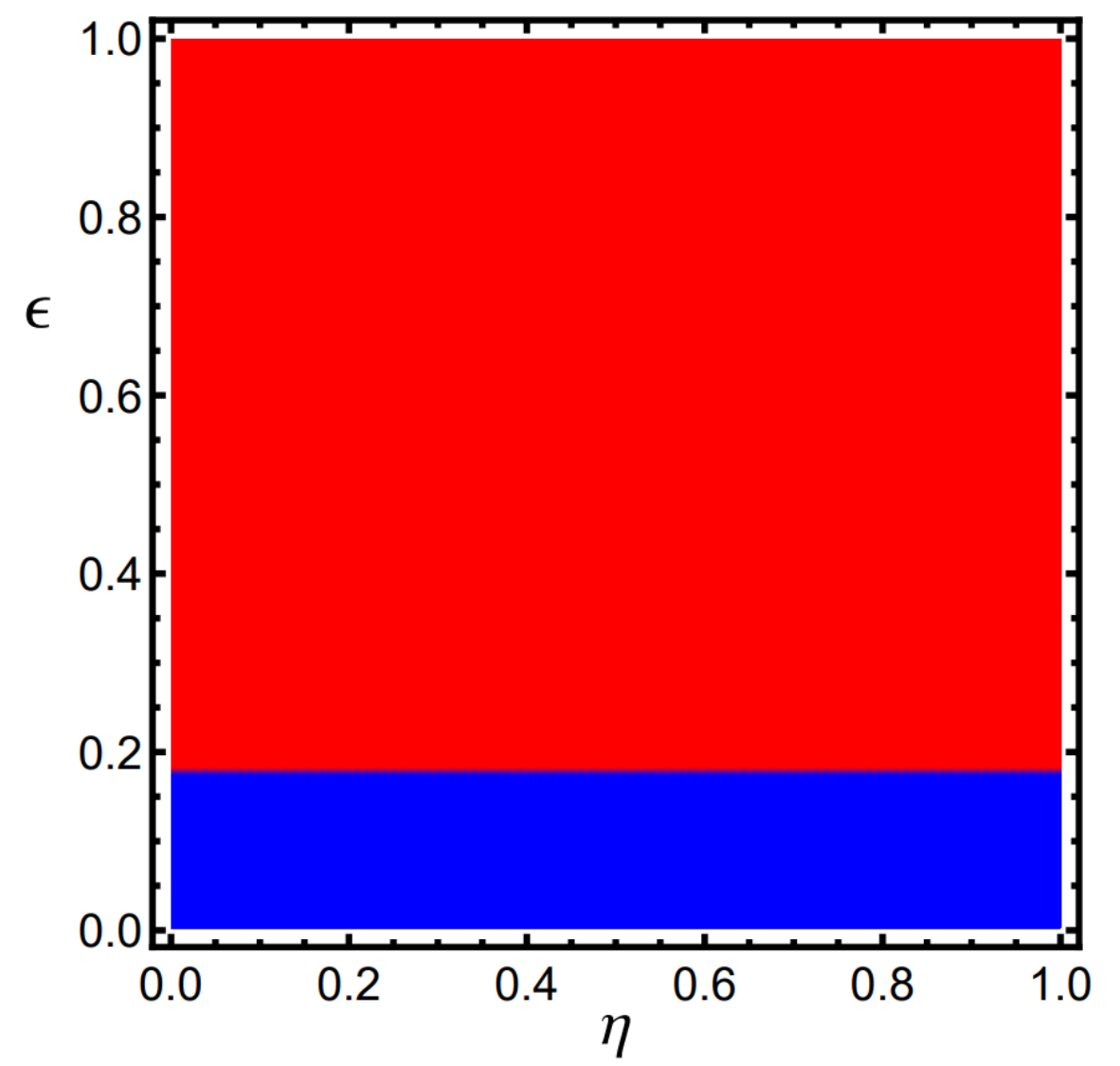}
	\caption{}
	\label{fig:dsQ2}
	\end{subfigure}
    \hfill
	\begin{subfigure}[b]{0.3\textwidth}
		\includegraphics[height=5cm]{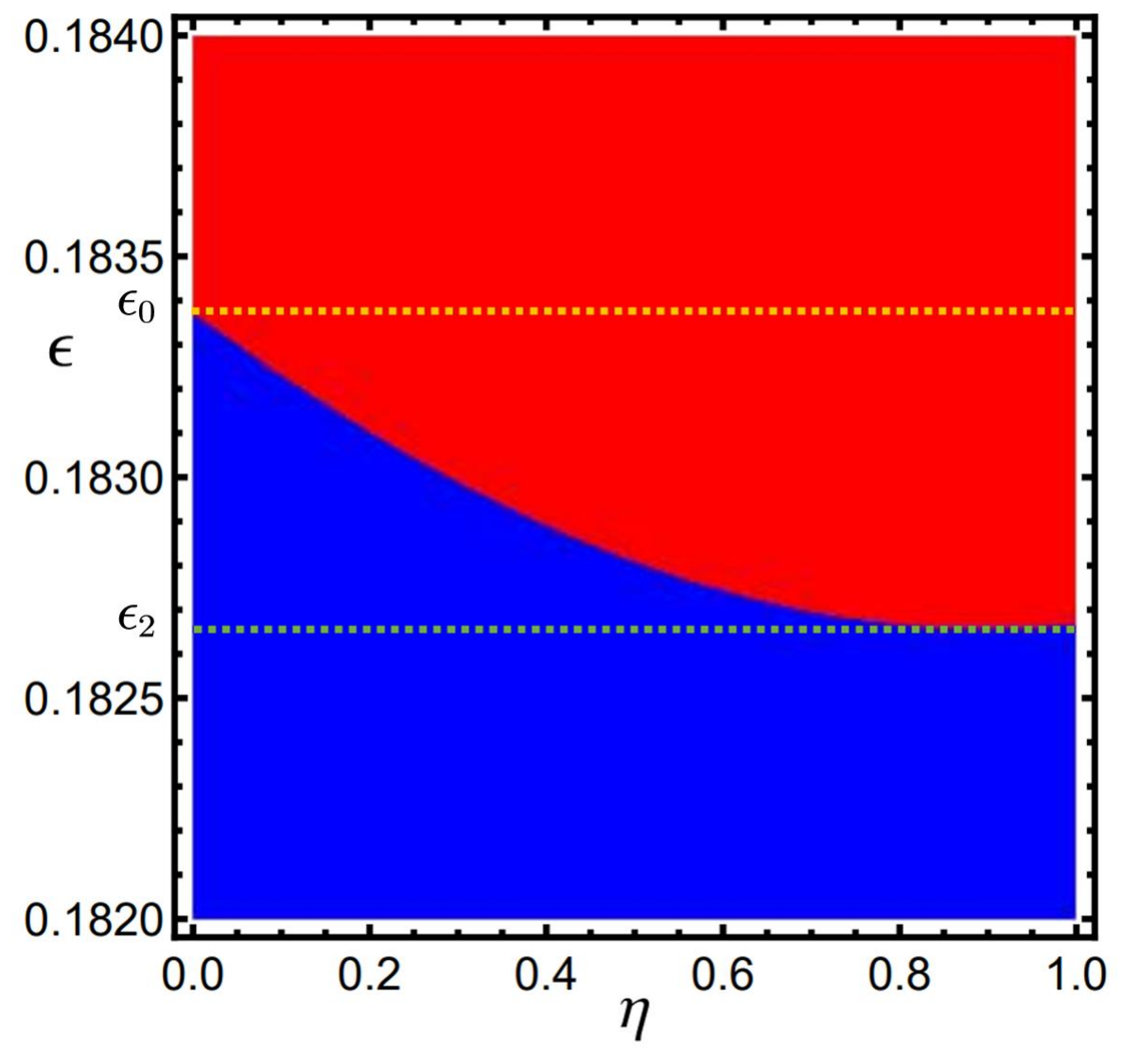}
		\caption{}
		\label{fig:dsQ}
	\end{subfigure}
	\hfill
	\begin{subfigure}[b]{0.3\textwidth}
		\includegraphics[height=5cm]{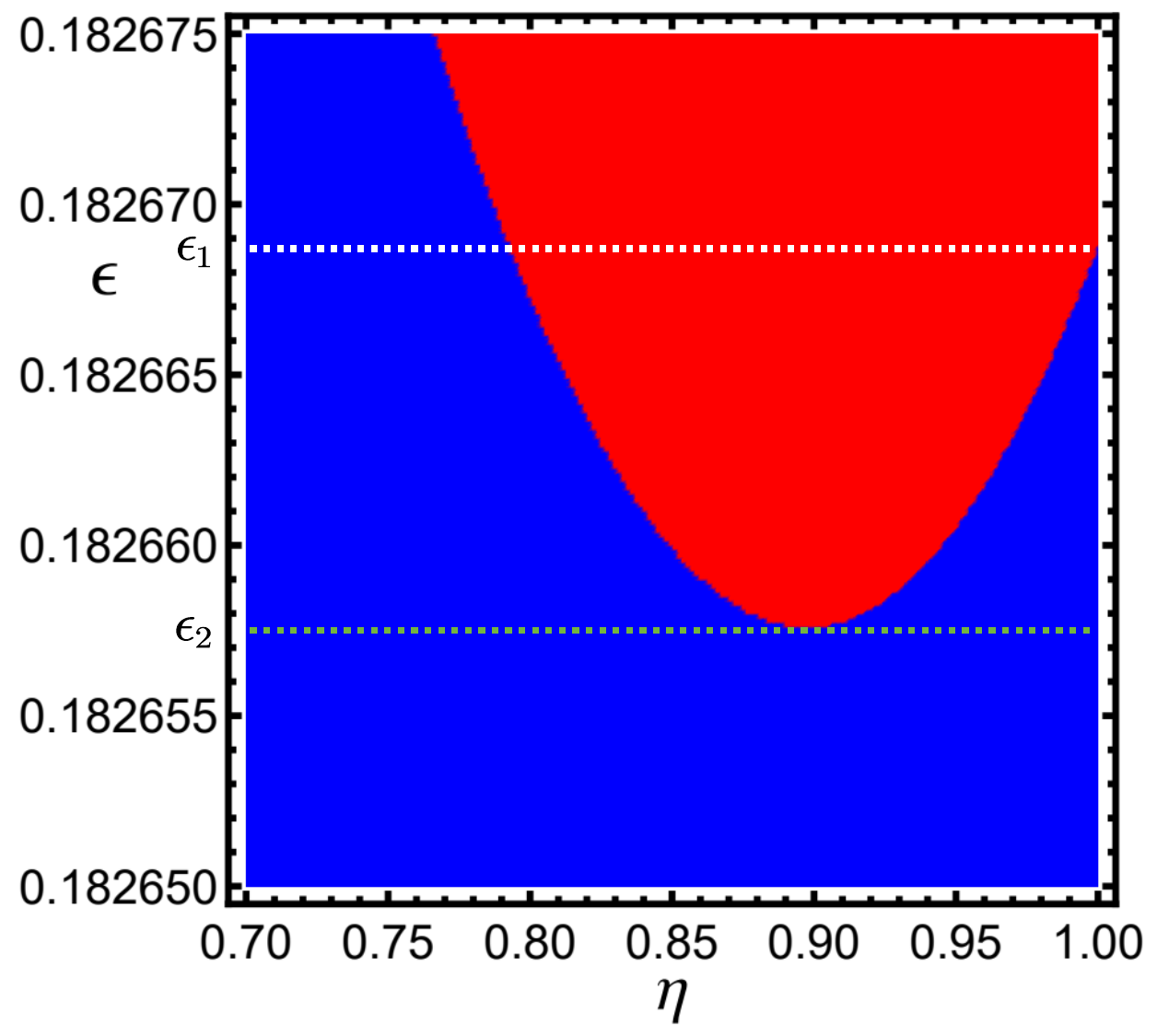}
		\caption{}
		\label{fig:dsQ1}
	\end{subfigure}
	\caption{
   (a) The sign of the first-order partial derivative of the angular separation $s$ with respect to the charge parameter $\eta$, where positive values are shown in red and negative values in blue. 
   (b) An enlarged view of Fig. \ref{fig:dsQ2} in the region $\epsilon \in \left[ 0.182,0.184 \right]$. 
   (c) An enlarged view of Fig. \ref{fig:dsQ2} in the region $\epsilon \in \left[ 0.182650,0.182675 \right]$ and $\eta \in \left[ 0.7,1 \right]$. 
   The yellow, white, and green dashed lines correspond to the critical values $\epsilon_0 \approx  0.1833742$, $\epsilon_1 \approx 0.1826688$, and $\epsilon_2 \approx 0.1826576$, respectively. 
	}\label{fig:dsQ_1}
\end{figure}

\begin{figure}[htbp]
		\centering
	\begin{subfigure}[b]{0.5\textwidth}
	\includegraphics[height=6.5cm]{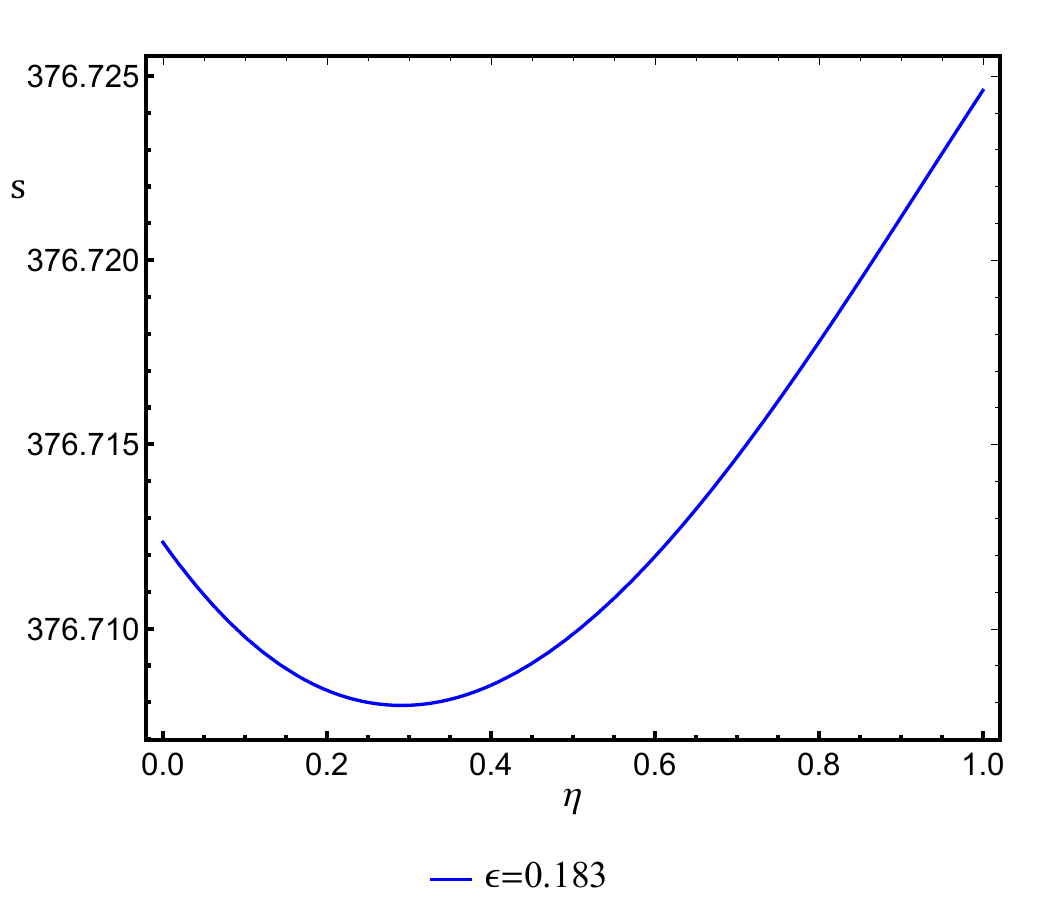}
		\caption{}
		\label{fig:s_1}
	\end{subfigure}
	\hfill
	\hspace{-1 cm}
	\begin{subfigure}[b]{0.5\textwidth}
		\includegraphics[height=6.5cm]{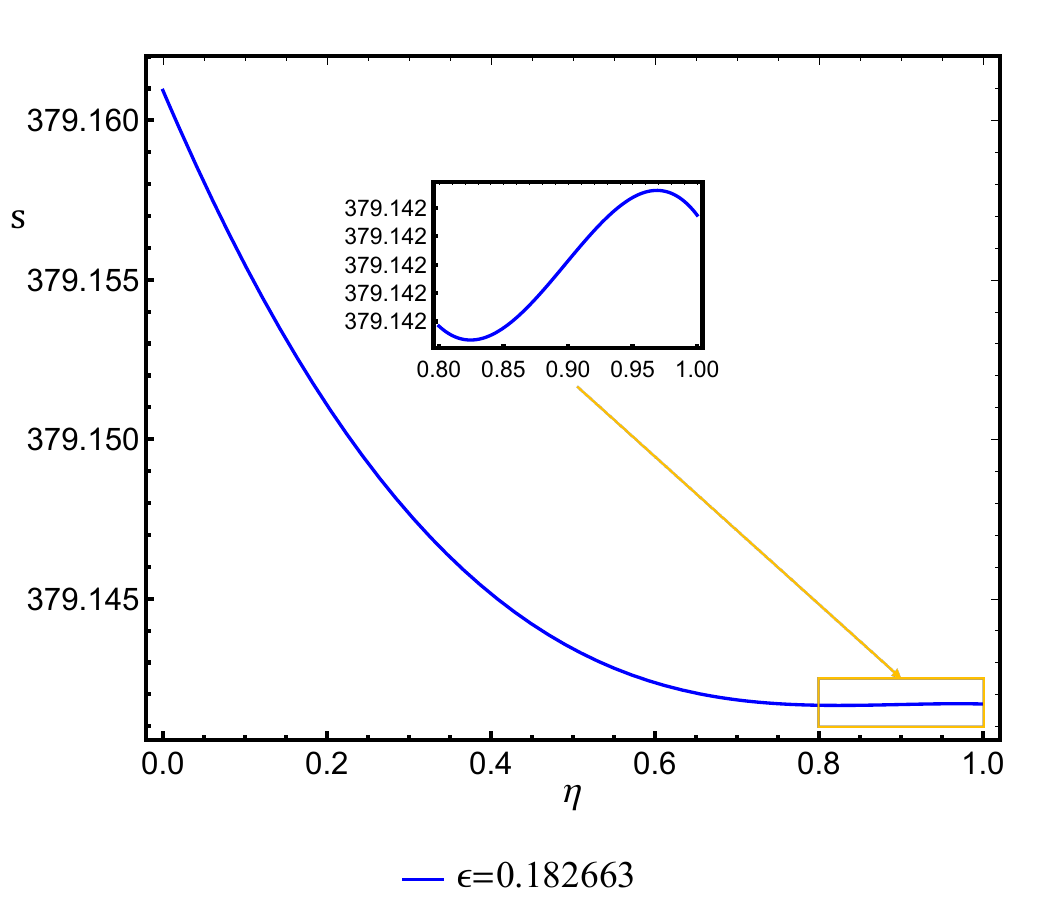}
		\caption{}
		\label{fig:s_2}
	\end{subfigure}
	\caption{ 
        (a) The dependence of angular separation $s$ on the charge parameter $\eta$ when $\epsilon=0.183$, where $\epsilon_1 < \epsilon < \epsilon_0$. 
        (b) The dependence of angular separation $s$ on the charge parameter $\eta$ when $\epsilon=0.182663$, where $\epsilon_2 < \epsilon < \epsilon_1$. 
        In both cases, the angular separation is measured in microarcsecond, and $D_\text{OL}=10^{10}M$. 
       }\label{fig:s1_s2}
\end{figure}

The analysis above focuses on the monotonicity of the angular separation in the small-charge ($\eta \to 0$) and large-charge ($\eta \to 1$) limits. To gain a more comprehensive understanding of its behavior for arbitrary black hole charge, we examine numerically the sign of the partial derivative of the angular separation $s$ with respect to the charge parameter $\eta$ as a function of the global monopole parameter $\epsilon$ and the charge parameter $\eta$, as shown in Fig. \ref{fig:dsQ_1}. Additionally, Figs. \ref{fig:dsQ_1}\subref{fig:dsQ} and \ref{fig:dsQ_1}\subref{fig:dsQ1} provide enlarged views of specific regions of Fig. \ref{fig:dsQ2}. 
The figures clearly reveal that, in addition to the two critical values $\epsilon_0$ and $\epsilon_1$ found in the analytical investigation, a third critical value $\epsilon_2 \approx 0.1826576$ emerges from the numerical analysis. 
These critical points divide the parameter space into four distinct regions that govern the monotonicity and behavior of the angular separation $s$   with respect to the charge parameter ($\eta$): 
\begin{enumerate}
\item When $\epsilon < \epsilon_2$, the partial derivative is always negative, indicating that the angular separation decreases monotonically with charge.
\item When $\epsilon_2 < \epsilon < \epsilon_1$, the angular separation initially decreases, then increases, and subsequently decreases again as the charge increases. This nonmonotonic behavior results in the angular separation exhibiting both a minimum and a maximum, as illustrated in Fig.~\ref{fig:s1_s2}.
\item When $\epsilon_1 < \epsilon < \epsilon_0$, the angular separation initially decreases and then increases as the charge increases. This again indicates the existence of a single minimum in the angular separation as a function of charge, as demonstrated in Fig.~\ref{fig:s1_s2}.
\item When $\epsilon > \epsilon_0$, the partial derivative is always positive, meaning that the angular separation increases monotonically with charge.
\end{enumerate}
These results reveal that  the presence of the global monopole introduces more complex behavior in the angular separation as a function of charge. In particular, the angular separation does not always increase with charge, as it does for the RN black hole. Instead, the behavior depends on the global monopole parameter $\epsilon$, with some regions exhibiting a minimum or maximum angular separation as charge varies.

\subsection{Flux ratio}

The flux ratio is given by Eq. \eqref{eq:flux} as
\bq\lb{eq:flux2}
	\mathcal{R}(\epsilon,\eta)=\frac{(e^{\bar{b}/\bar{a}}+e^{2\pi / \bar{a}})(e^{4\pi /\bar{a}}-1)}{e^{\bar{b}/\bar{a}}+e^{2\pi/\bar{a}}+e^{4\pi/ \bar{a}}},
\eq
where $\bar{a}$ and $\bar{b}$ denote $\bar{a}(\epsilon,\eta)$ and $\bar{b}(\epsilon,\eta)$, respectively. 
Direct analysis of the dependence of the flux ratio in Eq. \eqref{eq:flux2} on the charge and the global monopole parameter is complex. To gain a more intuitive understanding of the influence of the global monopole on the flux ratio, 
we focus on the case where the deficit angle caused by the global monopole is small, i.e., when $\epsilon$ is close to $1$. In this scenario, $e^{\frac{2\pi}{\bar{a}}}\gg e^{\frac{\bar{b}}{\bar{a}}}$, and thus Eq. \eqref{eq:flux2} can be approximated as
\bq\lb{eq:flux_3}
	\mathcal{R}(\epsilon,\eta)\approx e^{\frac{2\pi}{\bar{a}}}=\exp \left[ \frac{2\pi \sqrt{\epsilon \bar{r}_m^{2}-2\eta}}{\bar{r}_m} \right],  
\eq
where $\bar{r}_m$ represents $\bar{r}_m(\epsilon,\eta)$. 

First, we investigate the dependence of the flux ratio $\mathcal{R}$ on the charge parameter $\eta$ for a fixed global monopole parameter $\epsilon$.  From Eq. \eqref{eq:flux_3}, we obtain
\bq
\frac{\partial \mathcal{R}}{\partial \eta}=-\frac{6\pi}{\left[ \bar{r}_m K\left( \epsilon ,\eta \right) \right] ^{\frac{3}{2}}}\exp \left( 2\pi \sqrt{\frac{K\left( \epsilon ,\eta \right)}{\bar{r}_m}} \right) <0 ,
\eq
where $K\left( \epsilon,\eta \right) =\sqrt{9-8\epsilon \eta}$. 
Therefore, the flux ratio $\mathcal{R}$ decreases monotonically as the charge increases when the deficit solid angle induced by the global monopole is small (i.e., when $\epsilon$ is close to 1). 
Furthermore, we numerically investigate the partial derivative of the flux ratio $\mathcal{R}$ \eqref{eq:flux2} with respect to the charge parameter $\eta$, which is always negative in the region $\epsilon\in[0,1]$ and $\eta\in[0,1]$. This indicates that the flux ratio $\mathcal{R}$ decreases monotonically with $\eta$, irrespective of the value of the global monopole parameter $\epsilon$.

Next, we analyze how the flux ratio $\mathcal{R}$ varies with the global monopole parameter $\epsilon$ for a fixed charge. For $\epsilon$ close to $1$, Eq. \eqref{eq:flux_3} can be expanded as
\bq \lb{eq:fe}
\mathcal{R}=\exp \left[ 2\pi \sqrt{\frac{2W(\eta)}{G\left( \eta \right)}} \right] +\frac{2\sqrt{2}\pi I\left( \eta \right)}{G^{\frac{3}{2}}\left( \eta \right) W(\eta)}\exp \left[ 2\pi \sqrt{\frac{2W(\eta)}{G\left( \eta \right)}} \right] \left( 1-\epsilon \right) +\mathcal{O}\left(( 1-\epsilon)^2 \right) ,
\eq
where $W(\eta)=\sqrt{9-8\eta}$, $G\left( \eta \right) =3+\sqrt{9-8\eta}$, and 
\bq
I\left( \eta \right) =-32\eta ^2+6\left( 5\sqrt{9-8\eta}+21 \right) \eta-27\left( \sqrt{9-8\eta}+3 \right).
\eq
The sign of $I(\eta)$ determines whether the flux ratio $\mathcal{R}$ increases or decreases with the global monopole parameter $\epsilon$. From the above equations, we find a critical value $\eta_0=\frac{27}{32}$ (or equivalently, a critical charge $Q_0=\frac{3\sqrt{6}}{8}M$) such that $I(\eta_0)=0$. Specifically, the flux ratio $\mathcal{R}$ increases with $\epsilon$ (i.e., decreases with the deficit angle) when $\eta<\eta_0$, whereas $\mathcal{R}$ decreases as $\epsilon$ increases (i.e., as the deficit angle decreases) when $\eta>\eta_0$. 
This behavior is in sharp contrast to the case without charge, where the flux ratio increases monotonically with the global monopole parameter, or as the deficit angle decreases, as demonstrated numerically in Ref. \cite{Cheng:2010nd}.

\begin{figure}[htbp]
\includegraphics[height=6.5cm]{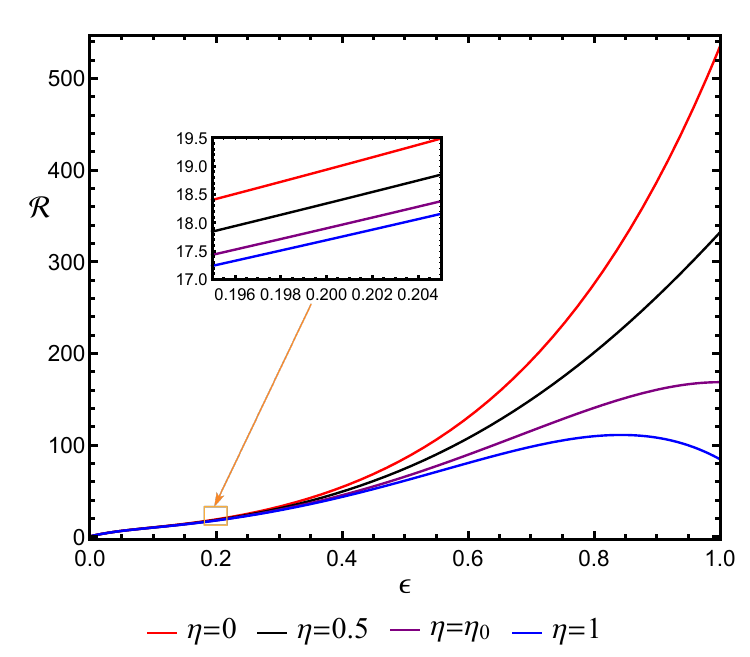}
\caption{The dependence of flux ratio $\mathcal{R}$ on the deficit angle characterized by $\epsilon$ for fixed charge. The solid red, black, purple, and blue lines correspond to charge parameter $\eta=Q^2/M^2$ with values of $0$, $0.5$, $\eta_0$, and $1$, respectively. The inset shows an enlargement of the interval $\epsilon\in [0.195, 0.205]$.
}\label{fig_flux}
\end{figure}

We now numerically analyze the behavior of the flux ratio $\mathcal{R}$ over a broader parameter range. 
As shown in Fig.~\ref{dR}, the critical value $\eta_0=\frac{27}{32}\approx 0.84$  (indicated by the yellow dashed line) divides the parameter space into two distinct regions. 
When the charge parameter $\eta$ is less than $\eta_0$,  the flux ratio $\mathcal{R}$ decreases monotonically with the global monopole parameter $\epsilon$. In contrast, when $\eta$ exceeds $\eta_0$, the flux ratio $\mathcal{R}$ initially increases and then decreases as $\epsilon$ increases.
To illustrate these behaviors explicitly, we show the flux ratio as a function of the global monopole parameter $\epsilon$ for a fixed charge in Fig. \ref{fig_flux}. 
From this figure, we observe that when the charge satisfies $\eta\leq\frac{27}{32}$ (i.e., the red and black lines), the flux ratio $\mathcal{R}$ increases as the global monopole parameter $\epsilon$ increases (i.e., as the deficit angle decreases). 
However, when $\eta>\eta_0$ (i.e., the blue line), the flux ratio $\mathcal{R}$ initially increases and then decreases. 
Furthermore, it can also be seen from Fig. \ref{fig_flux} that, when the global monopole parameter $\epsilon$ is fixed, the flux ratio $\mathcal{R}$ decreases as the charge (i.e., the parameter $\eta$) increases.

\begin{figure}[h!]
    \includegraphics[height=6.5cm]{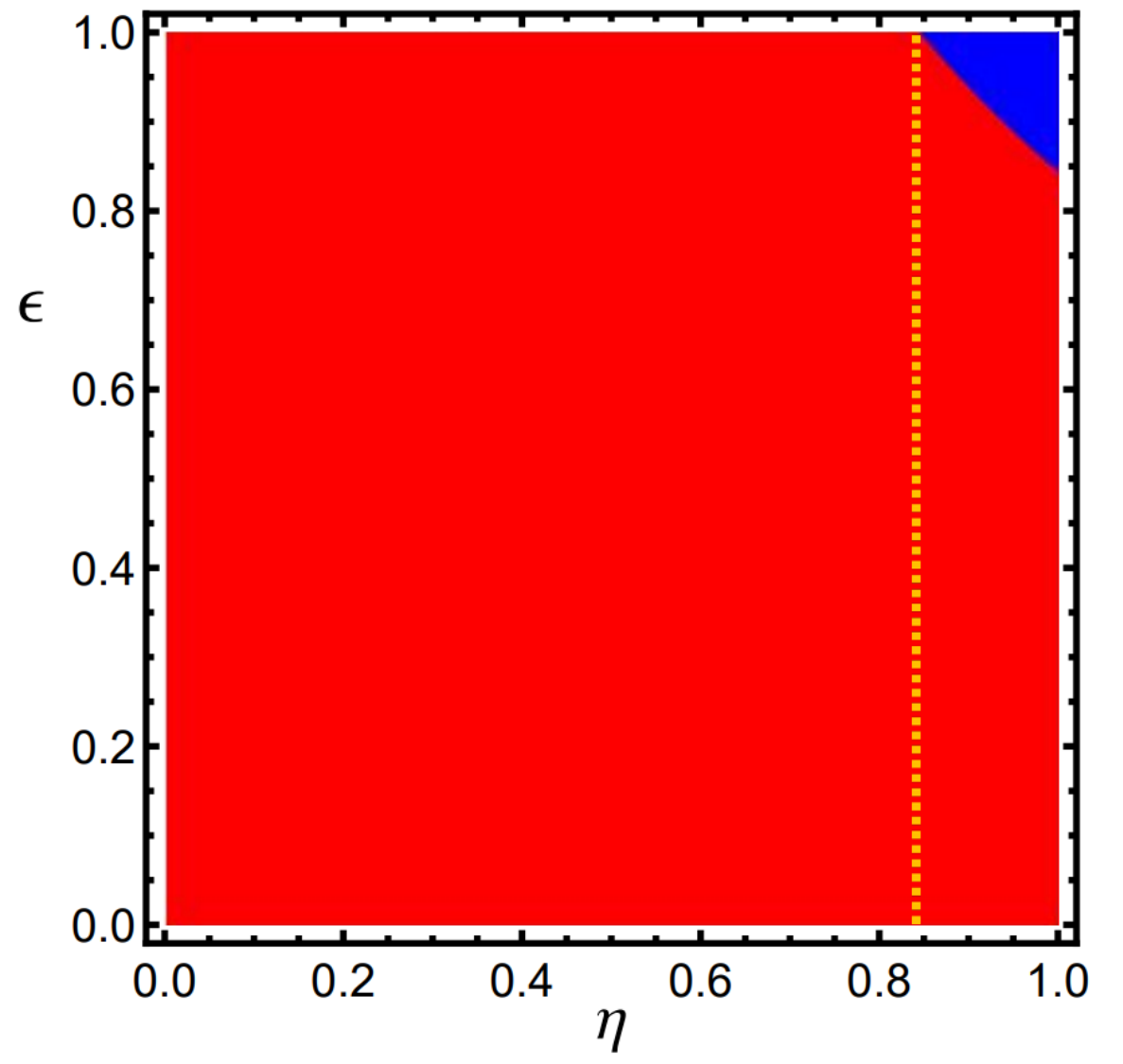}
	\caption{
        The sign of the first-order partial derivative of the flux ratio $\mathcal{R}$ with respect to the global monopole parameter $\epsilon$, where positive values are shown in red and negative values in blue. The yellow line represents the critical charge paremeter $\eta_0$.
    }\lb{dR}
\end{figure}

\subsection{Magnification of the first image}

The magnification of the first image,  denoted as $\mu _1$ can be obtained from Eq. (\ref{eq:mu}) as follows 
\bq\lb{eq:mu_1}
\mu _1(\epsilon,\eta)=e_1\frac{b_{m}^{2}\left( 1+e_1 \right) D_{\text{OS}}}{\bar{a}\beta D_{\text{OL}}^{2}D_{\text{LS}}},
\eq
where $\bar{a}$ and $b_m$ represent $\bar{a}(\epsilon,\eta)$ and $b_m(\epsilon,\eta)$, respectively, and $e_1=e^{\frac{\bar{b}-2\pi}{\bar{a}}}$.

\begin{figure}[htbp]
	\centering
    \begin{subfigure}[b]{0.5\textwidth}
	\includegraphics[height=6.5cm]{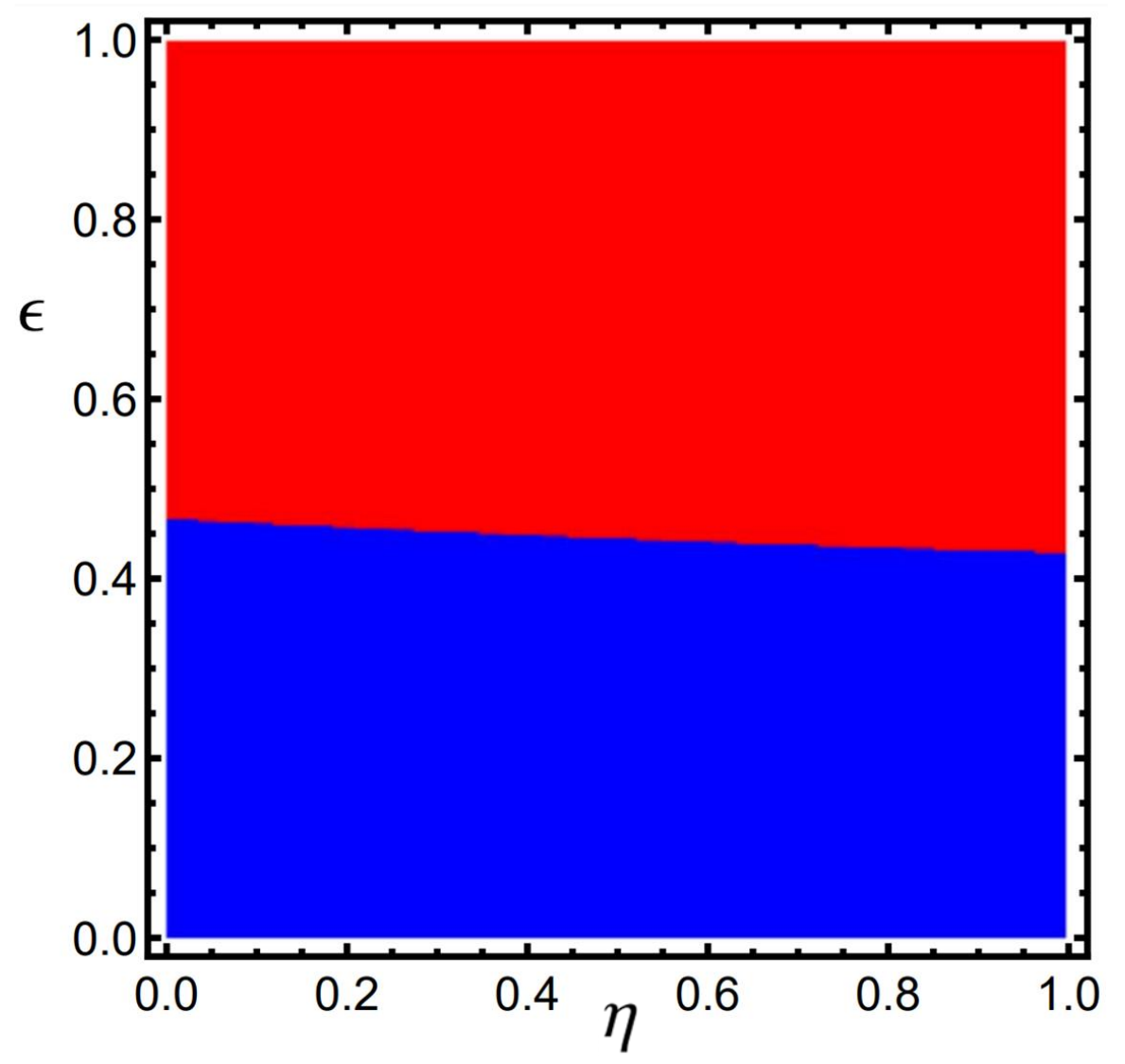}
	\caption{}
	\label{fig:dmuQ}
	\end{subfigure}
    \hfill
	\hspace{-1 cm}
	\begin{subfigure}[b]{0.5\textwidth}
		\includegraphics[height=6.5cm]{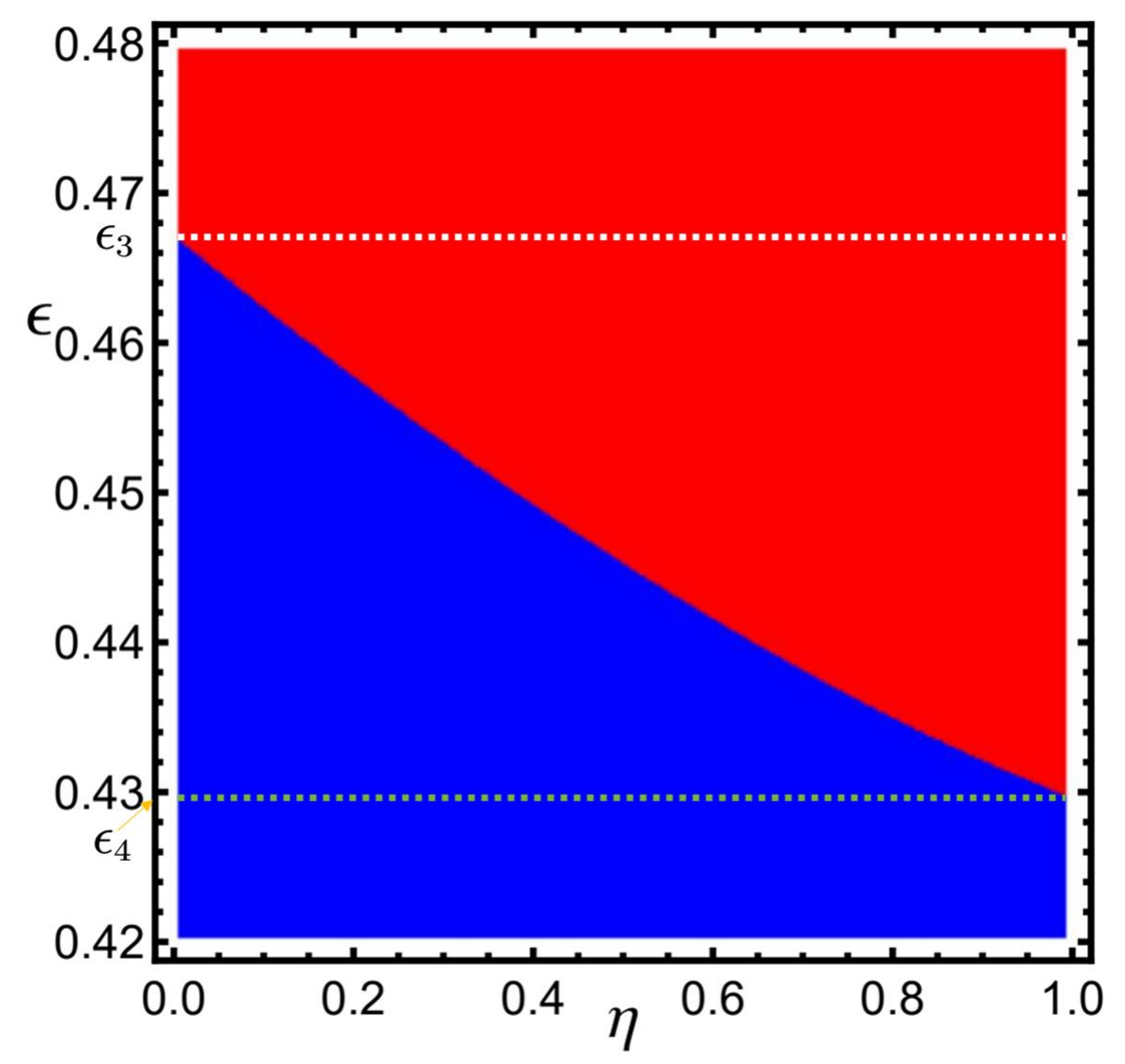}
		\caption{}
		\label{fig:dmuQ1}
	\end{subfigure}
	\caption{(a) The sign of the first-order partial derivative of the magnification $\mu_1$ of the first image with respect to the charge parameter $\eta$, where positive values are shown in red and negative values in blue. 
   (b) An enlarged view of Fig. \ref{fig:dmuQ} in the region $\epsilon \in \left[ 0.42,0.48 \right]$. 
   The white and green dashed lines correspond to the critical values $\epsilon_3 \approx  0.4674$ and $\epsilon_4 \approx 0.4295$, respectively. 
	}\label{fig:dmuQ_1}
\end{figure}

\begin{figure}[htbp]
\includegraphics[height=6.5cm]{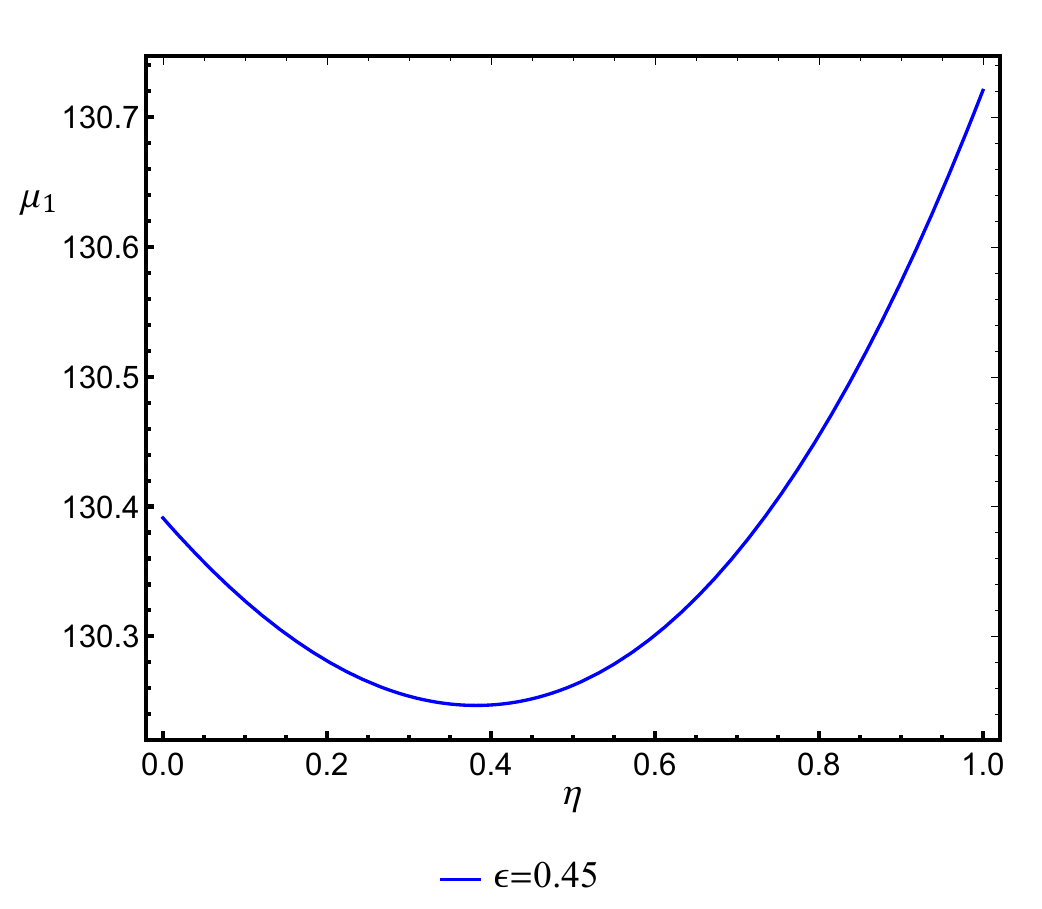}
\caption{The magnification $\mu_1$ of the first image as a function of the charge parameter $\eta$ for a fixed global monopole parameter $\epsilon$. The actual magnification is given by $\mu_1\times 10^{-20}$. Here $\epsilon=0.45$, and $D_\text{OL}=D_\text{LS}=10^{10}M$.}\label{fig:mu}
\end{figure}

Direct analytical examination of the behavior of $\mu_1$ is challenging, so we perform a numerical analysis. The sign of the first-order partial derivative of $\mu_1$ with respect to the charge parameter $\eta$ is presented in Fig. \ref{fig:dmuQ_1}, with Fig. \ref{fig:dmuQ1} providing an enlarged view of specific regions from Fig. \ref{fig:dmuQ}. These figures reveal the existence of two critical values $\epsilon_3 \approx 0.4674$ and  $\epsilon_4 \approx 0.4295$, which divide the parameter space into three distinct regions, each exhibiting different monotonic behaviors of the magnification $\mu_1$ with respect to the charge parameter $\eta$. 
\begin{enumerate}
\item When $\epsilon < \epsilon_4$, the partial derivative of the magnification with respect to charge is always negative, indicating that the magnification decreases monotonically as the charge increases. 
\item When $\epsilon_4 < \epsilon < \epsilon_3$, the magnification initially decreases and then increases as the charge increases, indicating the existence of a single minimum in the magnification as a function of charge, as demonstrated in Fig. \ref{fig:mu}. 
\item When $\epsilon > \epsilon_3$, the partial derivative is always positive, meaning that the magnification increases monotonically with charge. 
\end{enumerate}
Additionally, direct numerical calculations show that the partial derivative of $\mu_1$ with respect to the global monopole parameter $\epsilon$ is always negative, suggesting that the magnification decreases as $\epsilon$ increases.

\subsection{Size of the first Einstein ring}

\begin{figure}[htbp]
\centering
\includegraphics[height=6.5cm]{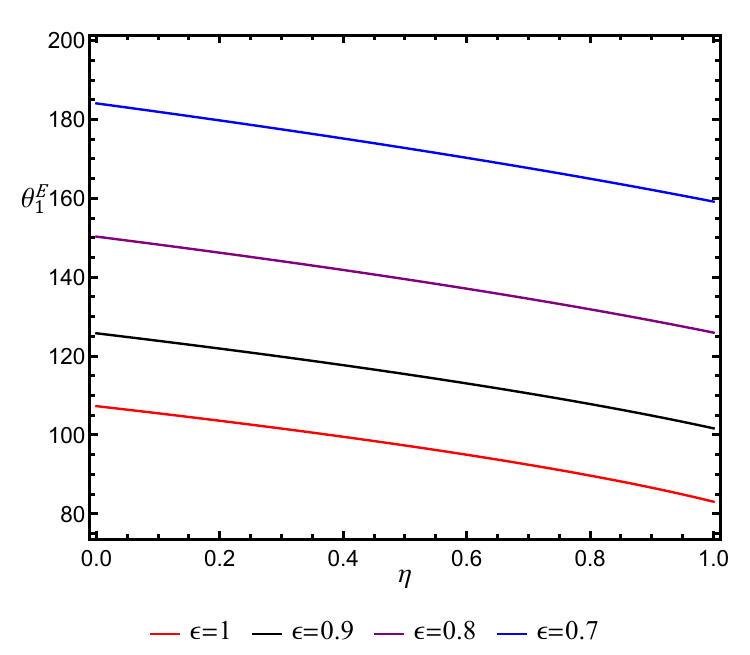}
\caption{Angular radius of the first Einstein ring $\theta_1^E$ as a function of the charge parameter $\eta$, for various values of the global monopole parameter $\epsilon$. The red, black, purple, and blue curves correspond to $\epsilon = 1$, $0.9$, $0.8$, and $0.7$, respectively. The angular radius is measured in microarcseconds. The lens-observer distance is set to $D_\text{OL} = 10^{10}M$.}
\label{fig:ring}
\end{figure}

The angular radius of the first Einstein ring can be well approximated by $\theta^0_1$, as given in Eq. \eqref{theta-0-n}. Substituting  the gravitational lensing coefficients $\bar{a},\bar{b}$ from  Eqs. \eqref{eq:abar_rad} and \eqref{eq:bbar_rad}, along with the critical impact parameter $b_m$ from Eq. \eqref{eq:b_m}, into Eq. \eqref{theta-0-n}, we obtain the explicit expression
\bq
\theta _{1}^{E}\left( \epsilon ,\eta \right) =\frac{M\bar{r}_{m}^{2}}{D_{OL}\sqrt{\epsilon \bar{r}_{m}^{2}-2\bar{r}_m+\eta}}\left\{ 1+\left[ \frac{8\left( 3\bar{r}_m-4\eta \right) ^3}{\bar{r}_{m}^{2}\left( \bar{r}_m-\eta \right) ^2}\left( 2\sqrt{\bar{r}_m-\eta}-\sqrt{3\bar{r}_m-4\eta} \right) ^2 \right] \exp \left( \frac{-3\pi \sqrt{\epsilon \bar{r}_{m}^{2}-2\eta}}{\bar{r}_m} \right) \right\},
\eq
where $\bar{r}_m$ is $\bar{r}_m(\epsilon,\eta)$. Due to the complexity of this expression, analytical analysis is intractable; therefore, we perform a numerical investigation.
Using the same method as before, we compute the partial derivatives of $\theta_1^E$ with respect to $\epsilon$ and $\eta$  and find both to be negative. This indicates that the radius of the first Einstein ring  $\theta_1^E$ always increases as the deficit solid angle $\Delta$ due to the global monopole grows (i.e., as $\epsilon$ decreases), irrespective of the charge of the black hole. Conversely, $\theta_1^E$ always decreases with increasing charge, regardless of the presence of a global monopole.
To visualize these behaviors,   Fig.~\ref{fig:ring} plots $\theta_1^E$ as a function of the  black hole charge parameter $\eta$ for several fixed values of $\epsilon$.  Each curve shows a monotonic decrease in $\theta_1^E$ with increasing $\eta$. Additionally, comparisons across curves reveal that a larger deficit angle (i.e., smaller $\epsilon$) leads to a larger Einstein ring.

\subsection{Time delay between the first and second images}

\begin{figure}[htbp]
\centering
\includegraphics[height=6.5cm]{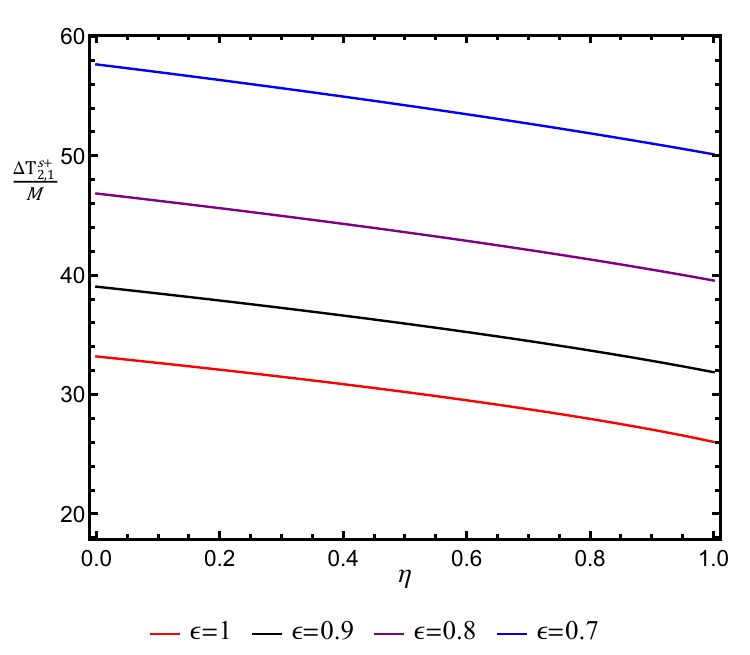}
\caption{Dimensionless time delay $\Delta T_{2,1}^{s+}/M$ between the first and second relativistic images as a function of the charge parameter $\eta$, for several fixed values of the global monopole parameter $\epsilon$. The solid red, black, purple, and blue lines correspond to $\epsilon = 1$, $0.9$, $0.8$, and $0.7$, respectively. The lens-observer distance is set to $D_\text{OL} = 10^{10}M$, with $D_\text{LS} = D_\text{OL} = D_\text{OS}/2$.}
\label{fig:time}
\end{figure}

According to Eq. \eqref{eq:timedelay}, the time delay between the first and second relativistic images, both formed on the same side of the lens and the source by photons traversing different paths, is given by 
  \bq\lb{eq:time}
    \Delta T_{2,1}^{s+}(\epsilon,\eta)=\frac{\bar{r}_{m}^{2}}{\sqrt{\epsilon \bar{r}_{m}^{2}-2\bar{r}_m+\eta}}\left[ 2\pi+2\sqrt{2}\left( \sqrt{e_1}-\sqrt{e_2} \right) + \frac{\sqrt{2}D_{\text{OS}}\beta}{\bar{a}D_{\text{LS}}}\left( \sqrt{e_1}-\sqrt{e_2} \right) \right].
    \eq
Here,  $\bar{r}_m = \bar{r}_m(\epsilon,\eta)$, $\bar{a} = \bar{a}(\epsilon,\eta)$, and $e_n$ is defined in Eq.~\eqref{eq:en}. 
Due to the complexity of this expression, we carry out a numerical investigation.
 Following the same methodology as in earlier sections,  we compute the partial derivatives of $\Delta T_{2,1}^{s+}$ with respect to $\epsilon$ and $\eta$. Our analysis reveals that both derivatives are consistently negative.
 This implies that the time delay increases with the  deficit angle $\Delta$ due to the global monopole (i.e., as $\epsilon$ decreases), regardless of whether the black hole is charged. Conversely, $\Delta T_{2,1}^{s+}$ decreases monotonically as the charge parameter $\eta$ increases, independently of the value of $\epsilon$.
To visualize these dependencies, Fig.~\ref{fig:time} displays $\Delta T_{2,1}^{s+}$ as a function of the charge parameter $\eta$ for various fixed values of the global monopole parameter $\epsilon$. Each curve clearly shows a decrease in time delay with increasing charge, while comparison across curves indicates that larger deficit angles (smaller $\epsilon$) yield longer time delays.

~

\section{Discussion}\lb{Discussion}

In this paper, we have investigated the gravitational lensing near a charged black hole with a global monopole in the strong field limit, focusing on the interplay between charge and the global monopole on lensing observables. 
Notably, the phenomena described here, such as the nonmonotonic behavior of the angular separation, are particularly intriguing when both charge and the global monopole's deficit angle are large, with critical values identified for both parameters (refer to the critical values for $\eta$ and $\epsilon$ in the last section). 
It is interesting to examine whether the effects of charge and the global monopole on lensing observables can be detected for reasonable values of the global monopole parameter and black hole charge in astrophysical settings. To this end, we consider black holes at the centers of three galaxies, i.e., the Milky Way, M87, and M104, as gravitational lenses. The relevant parameters are $M=4.3 \times 10^6 M_\odot$ and $D_\text{OL}=8.3\text{kpc}$ for Sgr A* \cite{GRAVITY:2023avo}, $M=6.5 \times 10^9 M_\odot$ and $D_\text{OL}=16.8\text{Mpc}$ for M87* \cite{EventHorizonTelescope:2019ggy}, and $M=1.0 \times 10^9 M_\odot$ and $D_\text{OL}=9.55\text{Mpc}$ for M104* \cite{1996ApJ,McQuinn2016THEDT}. 
Additionally, we assume that the lens is located between the light source and the observer, such that $D_\text{LS}=D_\text{OL}=D_\text{OS}/2$. The results for the angular separation $s$, the flux ratio $\mathcal{R}$, the magnification of the first image $\mu_1$, the angular radius of the first Einstein ring $\theta_{1}^{E}$, and the time delay between the first and second images $\Delta T_{2,1}^{s+}$ are presented in  Tables. \ref{table:s}-\ref{table:DeltaT}. These numerical results indicate that, for astrophysical black holes, the influence of charge and global monopoles on lensing observables is negligible. 
Even with exaggerated charge and deficit angle due to the global monopole ($\eta=0.01$ and $\Delta=4\pi\times 10^{-3}$, or equivalently $\epsilon=1-\frac{\Delta}{4\pi}=0.999$), the results remain nearly indistinguishable from those of Schwarzschild black holes.

In fact, astrophysical black holes are expected to be nearly neutral, as any significant charge would likely be neutralized by surrounding matter. Furthermore, the deficit angle caused by a global monopole is expected to be extremely small. In typical grand unification scenarios, the parameter $\gamma$ in the metric \eqref{eq:line} characterizing the global monopole is on the order of $\sim 10^{16}$ GeV \cite{Barriola:1989hx}, leading to a deficit angle $\Delta=32\pi^2\gamma^2\sim 4\pi\times 10^{-5}$ rad, which is negligibly small.  
Consequently, the rich gravitational lensing phenomena described in this paper are unlikely to be observed in typical astrophysical black holes.

\begin{table}
	\centering
	\begin{tabular}{|c|c|c|c|c|}
		\hline
		---& Schwarzschild &  \makecell{Schwarzschild with\\ global monopole}  & RN & \makecell{RN with\\ global monopole} \\ \hline
		Sgr A*& 0.0333 & 0.0335 & 0.0335 & 0.0337 \\ \hline
		M87*& 0.0248 & 0.0250 & 0.0250 & 0.0251 \\ \hline
		M104*& 0.0067 & 0.0068 & 0.0068 & 0.0068 \\ \hline
	\end{tabular}
	\caption{The angular separation $s$ (in  microarcsecond) between the first image and the other images for different black holes as lenses.     The charge and global monopole parameters are taken as $\eta=0.01$ and $\epsilon=0.999$, respectively. The same parameters apply to all following tables. } \label{table:s}
\end{table}

\begin{table}
	\centering
	\begin{tabular}{|c|c|c|c|c|}
		\hline
		---& Schwarzschild &  \makecell{Schwarzschild with\\ global monopole}  & RN & \makecell{RN with\\ global monopole} \\ \hline
		Sgr A*& 535.16 & 533.48 & 531.42 & 529.75 \\ \hline
		M87*& 535.16 & 533.48 & 531.42 & 529.75 \\ \hline
		M104*& 535.16 & 533.48 & 531.42 & 529.75 \\ \hline
	\end{tabular}
	\caption{The flux ratio $\mathcal{R}$ for different black holes as lenses.} \label{table:R}
\end{table}

\begin{table}
	\centering
	\begin{tabular}{|c|c|c|c|c|}
		\hline
		---& Schwarzschild &  \makecell{Schwarzschild with\\ global monopole}  & RN & \makecell{RN with\\ global monopole} \\ \hline
		Sgr A*& 4.7688 & 4.8034 & 4.7843 & 4.8189 \\ \hline
		M87*& 2.6597 & 2.6790 & 2.6684 & 2.6877 \\ \hline
		M104*& 0.1948 & 0.1962 & 0.1954 & 0.1969 \\ \hline
	\end{tabular}
	\caption{The magnification of the first image $\mu_1\times 10^{22}$ for different black holes as lenses.} \label{table:mu1}
\end{table}

\begin{table}
	\centering
	\begin{tabular}{|c|c|c|c|c|}
		\hline
		---& Schwarzschild &  \makecell{Schwarzschild with\\ global monopole}  & RN & \makecell{RN with\\ global monopole} \\ \hline
		Sgr A*& 26.613 & 26.653 & 26.569 & 26.609 \\ \hline
		M87*& 19.875 & 19.905 & 19.842 & 19.872 \\ \hline
		M104*& 5.379 & 5.387 & 5.370 & 5.378 \\ \hline
	\end{tabular}
	\caption{The angular radius of the first Einstein ring $\theta_{1}^{E}$ (in microarcsecond) for different black holes as lenses.} \label{table:ring}
\end{table}

\begin{table}
	\centering
	\begin{tabular}{|c|c|c|c|c|}
		\hline
		---& Schwarzschild &  \makecell{Schwarzschild with\\ global monopole}  & RN & \makecell{RN with\\ global monopole} \\ \hline
		Sgr A*& 0.1953 & 0.1956 & 0.1950 & 0.1953 \\ \hline
		M87*& 295.2550 & 295.7090 & 294.7790 & 295.2330 \\ \hline
		M104*& 45.4239 & 45.4938 & 45.3507 & 45.4205 \\ \hline
	\end{tabular}
	\caption{The time delay between the first and second images $\Delta T_{2,1}^{s+}$ (in hour) for different black holes as lenses.} \label{table:DeltaT}
\end{table}

Nevertheless, there is growing interest in analog gravity systems \cite{Barcelo:2005fc} as platforms for studying gravitational effects, particularly gravitational lensing \cite{Fischer:2001jz,sheng2013trapping}. These systems, which use acoustic, optical, or other artificial materials to mimic the behavior of spacetime, allow for more flexibility in tuning parameters such as charge and deficit angle, which may be too small in natural systems. 
In particular, the theory of transformation optics \cite{Leonhardt2006OpticalCM,pendry2006controlling} establishes a correspondence between the metric tensor and the permittivity and permeability tensors of materials, enabling the design of optical metamaterials to mimic astronomical phenomena. For instance, gravitational lensing effects have been experimentally visualized in such systems \cite{sheng2013trapping}. In such optical metamaterials, the effective charge and deficit angle are not constrained to be small. In particular, by using an artificial waveguide bounded with rotational metasurface, the nontrivial effects of a spacetime topological defect have been experimentally emulated, and photon deflection in the topological waveguide has been observed \cite{Sheng:2018jmy}. In this context, the phenomena predicted here may potentially be observed in analog systems.

\section{Conclusion}\lb{conclusion}

In this paper, we have investigated the phenomenon of gravitational lensing near a charged black hole with a global monopole,  with particular focus on the effects of the black hole charge and the global monopole on key lensing observables.  We have derived the coefficients $\bar{a}$ and $\bar{b}$ that characterize the deflection angle of light. Based on these results, we have investigated the effects of the global monopole and the black hole charge on key observables in gravitational lensing, including the angular separation, the flux ratio, the magnification, the angular radius of the Einstein ring,  and the time delay. 
The findings are summarized as follows:

For the angular separation $s$, we find that regardless of the value of the charge, $s$ always decreases as the deficit angle caused by the global monopole decreases. However, the behavior of 
$s$ as a function of the charge is highly dependent on the global monopole parameter. 
We identified three critical values of the global monopole parameter, which divide the parameter space into four distinct regions. 
In these regions, the angular separation $s$ can either increase or decrease monotonically with charge, or exhibit more complex behaviors such as an initial decrease followed by an increase, or a nonmonotonic behavior with both a maximum and a minimum. 
These behaviors contrast sharply with the monotonic increase in angular separation as a function of charge observed in the absence of a global monopole.

For the flux ratio $\mathcal{R}$, we find that irrespective of the global monopole parameter, $\mathcal{R}$ always decreases as the charge increases. However, the dependence of $\mathcal{R}$ on the global monopole parameter, or the deficit angle,  varies with the charge $Q$.   A critical charge $Q_0$ exists such that when the charge is smaller than this critical value, $\mathcal{R}$ increases as the deficit angle caused by the global monopole decreases. In contrast, when the charge exceeds the critical value, $\mathcal{R}$ initially increases and then decreases with decreasing deficit angle, exhibiting a maximum value. This behavior also differs from the monotonically decreasing flux ratio as the deficit angle increases in the absence of charge.

Similarly, for the magnification of the first relativistic image $\mu_1$, we find that irrespective of the charge, $\mu_1$ consistently  decreases as the global monopole parameter increases. However, the dependence of $\mu_1$ on the charge varies with the global monopole parameter. Specifically, two critical values of the global monopole parameter determine whether $\mu_1$ increases or decreases monotonically with charge, or whether it initially decreases and then increases.

These results demonstrate that in the strong field limit, the charge and global monopole interact in complex ways, and their combined effects on the lensing observables cannot be attributed to the independent contributions of each factor. The resulting lensing signatures provide a rich landscape for further exploration. In addition, we numerically analyzed the angular size of the first Einstein ring $\theta_1^E$, and the time delay between the first and second images $\Delta T_{2,1}^{s+}$ as functions of the charge and global monopole parameters. 
Our findings indicate that both observables decrease with increasing global monopole parameter $\epsilon$ and increasing charge.

While the charge and global monopole deficit angle are expected to be small in typical astrophysical black holes, these parameters are not constrained in analog gravity systems. Therefore, the phenomena predicted here, including the nonmonotonic behavior of the angular separation and flux ratio, may potentially be observed in optical metamaterials or other experimental setups designed to mimic black hole physics. These findings open up new avenues for testing gravitational lensing effects in laboratory-based systems, offering the possibility of future experimental verification of our theoretical predictions.

\section*{Acknowledgments}
		
This work was supported in part by the NSFC under Grant No. 12075084, and the innovative research group of Hunan Province under Grant No. 2024JJ1006.

\appendix

\section{DERIVATIVE OF ANGULAR SEPARATION WITH RESPECT TO $\eta$}\lb{ds}
\renewcommand{\theequation}
{\arabic{equation}}\setcounter{equation}{0}

\begin{figure}[h!]
	\includegraphics[height=6.5cm]{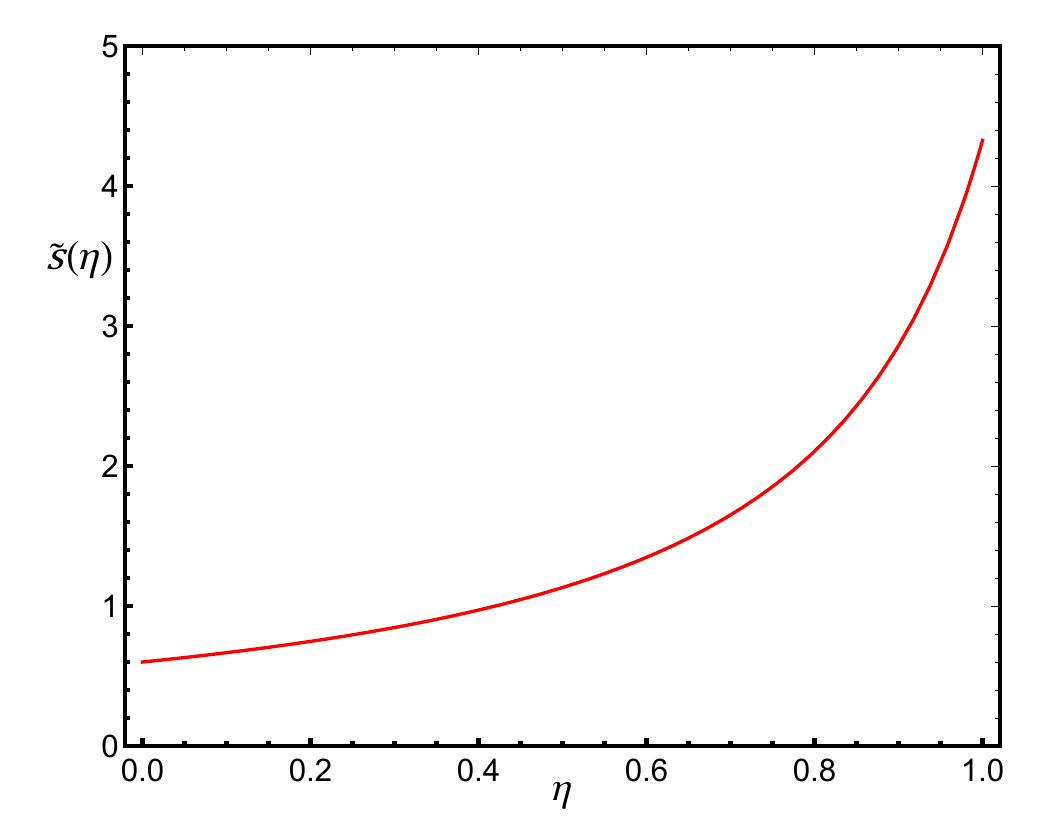}
	\caption{The behavior of $\tilde{s}(\eta)$ with $\eta$.
	}\label{fig_ds}
	\end{figure}

$s(1,\eta)$ is given by
\bq
s\left( 1,\eta \right) =\frac{M}{D_{\text{OL}}\sqrt{w^2-2w+\eta}}\left[ \frac{8\left( 3w-4\eta \right) ^3}{\left( w-\eta \right) ^2} \right] \left( 2\sqrt{w-\eta}-\sqrt{3w-4\eta} \right) ^2\exp \left( \frac{-3\pi \sqrt{w^2-2\eta}}{w} \right), 
\eq
where
\bq
w=\frac{3+\sqrt{9-8\eta}}{2}.
\eq
Taking the logarithmic derivative of $s(1,\eta)$, we obtain
\bq
\begin{split}
\tilde{s}(\eta) :=\frac{D_\text{OL}}{8M} \frac{ d\ln s(1,\eta)}{ d \eta}=&\frac{  \left(16 \eta +4w+4 \sqrt{w- \eta } \sqrt{3w-4 \eta}-33\right)}{2 (\eta -1) (8 \eta -9)}\\
&+\frac{\left(4 \sqrt{3w-4 \eta } \sqrt{w- \eta}-9\right) \left(w-3\right)}{2\eta (\eta -1) (8 \eta -9)}\\
&-\frac{18 \sqrt{2}\pi  \left[2 \left(2w+3\right) \eta -9w\right]}{w^{2} \left(6w-8 \eta\right)^{3/2} (2w-3)}
\end{split}
\eq
The trend of $\tilde{s}$ changing with $\eta$ is shown in the Fig. \ref{fig_ds}. Obviously, $\tilde{s}(\eta)$ is always positive.

	\section{EXPLICIT EXPRESSIONS FOR $H(\epsilon)$ AND $F(\epsilon)$ IN EQ. \eqref{eq:se2}}\lb{appendix}
		\renewcommand{\theequation}{\arabic{equation}}\setcounter{equation}{0}
	In this appendix, we provide the explicit expressions for $H(\epsilon)$ and $F(\epsilon)$ in Eq. \eqref{eq:se2}. For convenience, we define the following three functions:
    \bq
	h\left( \epsilon \right) =\sqrt{9-8\epsilon},
	\eq
	\bq	
	f_1\left( \epsilon \right) =3+\sqrt{9-8\epsilon},
	\eq	
    and
	\bq
	f_2\left( \epsilon \right) =\sqrt{f_1\left( \epsilon \right) -2\epsilon}.
	\eq	
	The function $H(\epsilon)$ is given by
	\bq
	H\left( \epsilon \right) =\frac{2\sqrt{2}}{f_2\left( \epsilon \right) ^5\epsilon ^{3/2}}\left[ f_1\left( \epsilon \right) h\left( \epsilon \right) \right] ^3\left[ \sqrt{f_1\left( \epsilon \right) h\left( \epsilon \right)}-2f_2\left( \epsilon \right) \right] ^2\exp \left[ 3\sqrt{2}\pi \sqrt{\frac{\epsilon h\left( \epsilon \right)}{f_1\left( \epsilon \right)}} \right], 
	\eq
	and the function $F(\epsilon)$ takes the form
	\bq
	F\left( \epsilon \right) =-\frac{2\left[ \sqrt{f_1\left( \epsilon \right) h\left( \epsilon \right)}-2f_2\left( \epsilon \right) \right]}{\left[ f_1\left( \epsilon \right) h\left( \epsilon \right) \right] ^{\frac{3}{2}}f_2\left( \epsilon \right) ^7\epsilon ^{7/2}}\left[ N_1\left( \epsilon \right) +N_2\left( \epsilon \right) \right] \exp \left[ 3\sqrt{2}\pi \sqrt{\frac{\epsilon h\left( \epsilon \right)}{f_1\left( \epsilon \right)}} \right], 
	\eq
	where
	\bq
N_1\left( \epsilon \right) =4\epsilon ^3h\left( \epsilon \right) ^3f_1\left( \epsilon \right) ^4f_2\left( \epsilon \right) \left[ \left( h\left( \epsilon \right) +2 \right) \sqrt{f_1\left( \epsilon \right) h\left( \epsilon \right)}-\left( 2h\left( \epsilon \right) +3 \right) f_2\left( \epsilon \right) \right], 
	\eq
	and
	\bq
	\begin{split}
N_2\left( \epsilon \right) =&36\pi \sqrt{2}\epsilon ^{7/2}\left[ \sqrt{f_1\left( \epsilon \right) h\left( \epsilon \right)}-2f_2\left( \epsilon \right) \right] \left[ 8\epsilon ^2-\left( 7h\left( \epsilon \right) +33 \right) \epsilon +9f_1\left( \epsilon \right) \right] \left[ f_1\left( \epsilon \right) h\left( \epsilon \right) \right] ^2\\
&+2\epsilon ^3\left[ \sqrt{f_1\left( \epsilon \right) h\left( \epsilon \right)}-2f_2\left( \epsilon \right) \right] \left[ -32\epsilon ^2+44\left( h\left( \epsilon \right) +6 \right) \epsilon -99f_1\left( \epsilon \right) \right] \left[ f_1\left( \epsilon \right) h\left( \epsilon \right) \right] ^{5/2}.
	\end{split}
	\eq

\begin{figure}[h!]
	\includegraphics[height=6.5cm]{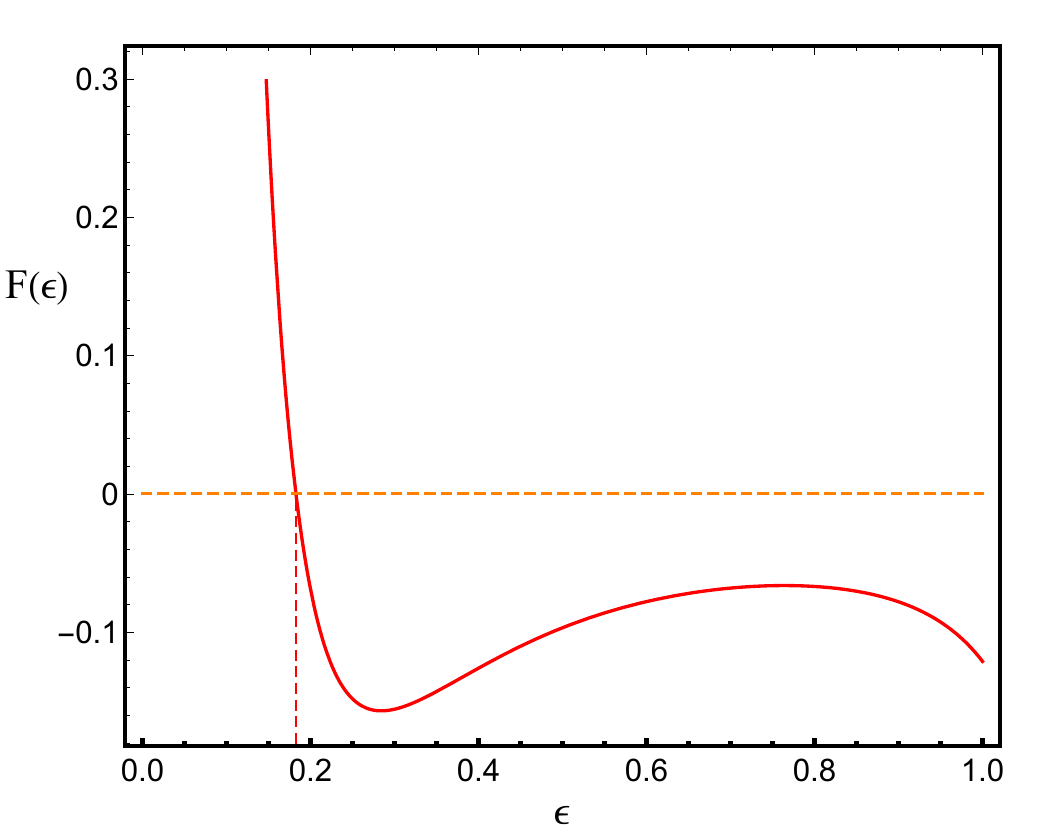}
	\caption{$F(\epsilon)$ as a function of the global monopole parameter $\epsilon$. 
    The abscissa of the zero point of the function is $\epsilon_1$.
	}\label{fig_F}
	\end{figure}

The numerical solution of $F(\epsilon_1)=0$ gives $\epsilon _1 \approx 0.1826688$. To illustrate the behavior of $F(\epsilon)$ around $\epsilon_1$, we present a plot of $F(\epsilon)$ in Fig. \ref{fig_F}. From Fig. \ref{fig_F}, it is evident that $F(\epsilon)>0$ when $\epsilon<\epsilon_1$, while $F(\epsilon)<0$ when $\epsilon>\epsilon_1$.

 \bibliographystyle{apsrev4-1} 
	\bibliography{RN}

	\end{document}